\documentclass[aps,prc,showpacs,twocolumn]{revtex4-1}

\usepackage{epsfig}
\usepackage{graphicx,amsmath}
\usepackage{color}
\newcommand\ba{\begin{eqnarray}}
\newcommand\ea{\end{eqnarray}}

\newcommand{\be}{\begin{equation}}
\newcommand{\ee}{\end{equation}}
\newcommand{\bas}{\begin{eqnarray*}}
\newcommand{\eas}{\end{eqnarray*}}
\begin{document}
\title{\bf \large Contribution to the study of baryonic spectroscopy, using baryon mass ratios}

\author{B. Tatischeff$^{1,2}$\\
$^{1}$Univ Paris-Sud, IPNO, UMR-8608, Orsay, F-91405\\
$^{2}$CNRS/IN2P3, Orsay, F-91405}
\thanks{tati@ipno.in2p3.fr}
 
\pacs{    }
\vspace*{1cm}
\begin{abstract}
The ratios between different baryonic species  masses are studied. The
result is used to tentatively predict some missing baryonic masses,
still not experimentally observed. 
\end{abstract}
\maketitle
\section{Introduction}
The study of fractal properties of hadronic spectroscopy \cite{bor1}, lead us to
observe that the distributions of $m_{n+1}/m_{n}$ mass ratios versus the rank, are almost the same among most species. Here $m_{n}$ is the $n^{th}$ mass of the given species at rank ``n''. This close behaviour was summarized in figure~34 of  \cite{bor1}, where all the first masses of Particle Data Group (PDG) \cite{pdg} were plotted after global translations of each species, in order to equalize the yrast mass of all species to the yrast charmed baryon mass. 
\begin{figure}[h]
\begin{center}
\hspace*{-3.mm}
\scalebox{0.95}[1.05]{
\includegraphics[bb=8 30 525 525,clip,scale=0.45]{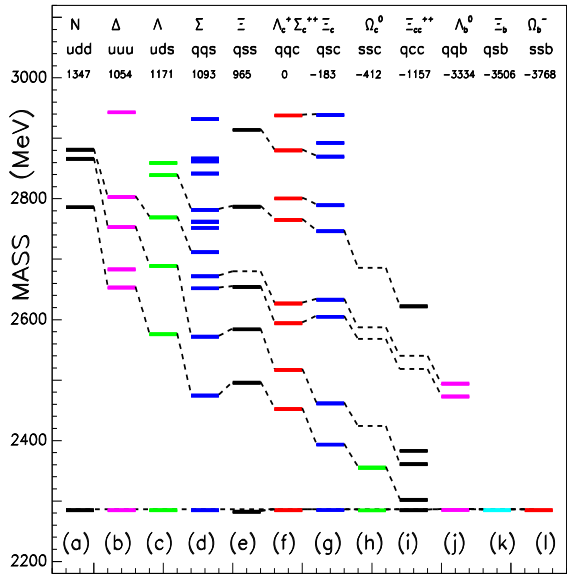}}
\caption{Comparison of all PDG baryon masses up to  M=2940~MeV, after global vertical translations of each species, in order to equalize the yrast mass of all species to the yrast charmed baryon mass (see text).}
\end{center}
\end{figure}

Although such comparison would be more justified after scaling, rather than translation, the translation allows to better visualize the regularity of the variation between excited masses of different baryonic species.
Figure~1 reproduces this figure~34 of  \cite{bor1}. A global vertical translations is done, in order to equalize the yrast mass of all species to the yrast charmed baryon mass. The amount of this mass translation is indicated at the top of the figure, below the quark content of each species.  For increasing yrast mass species, we observe that the gap between the yrast and the second mass decreases for all species except for $\Xi$ baryons.  We observe also that,
if we except the yrast masses, the distributions all baryonic species masses display close variations. 

In order to study this more quantitatively, we will show the ratios of the mass distributions of all baryonic species. We see from figure~1 that, concerning heavier species, very few baryonic masses have been observed  and no one in some cases. These species, $\Omega$ and all baryons containing a "b" quark, are not considered here. In all species, the masses are less defined, above an excitation of several hundred  MeV. It is also possible that at present, missing masses exist, which will be observed later. Therefore our study concerns only the first several hundred MeV excitation masses.
\section{Ratios between  different PDG baryonic species masses}
The PDG baryonic masses \cite{pdg} are reported (in black)  in table~1. 

 When  the masses change by a few MeV corresponding to different charges, their mean value is kept. For example the three
$\Sigma^{+,0,-}$ masses being respectively M = 1189.37~MeV, 1192.642~MeV, and 1197.449~MeV, the mean value M = 1193~MeV is used.
\subsection{Baryonic masses compared to $N^{*}$ masses}
Figures 2, 3,  4, 5, 6, and 7  show respectively the ratios of $\Delta$, $\Lambda$, $\Sigma$, $\Xi$, $\Lambda_{C}$, and $\Xi_{C}$ baryons over $N^{*}$. We observe that these ratios, represented by
full red circles, can generally be joined by one (or a few) horizontal line(s), up to rank 10, with two exceptions. The first exception concerns the first, fundamental (yrast) mass of every species. The second exception concerns the ratio of $\Xi / N^{*}$ masses (figure~5). In this figure the black full squares show ratios after introduction of still unobserved $\Xi$  baryons; this point will be discussed later. In this figure, the first four baryonic masses are the masses reported by PDG, therefore the same marks apply for full red circles and full black squares.  When more than $\approx$~10 data exist, the ratio in this range exhibits a second horizontal line slightly lower than the previous one.
This point will be discussed below.
\begin{figure}[h]
\begin{center}
\hspace*{-3.mm}
\vspace*{5.mm}
\scalebox{1}[0.9]{
\includegraphics[bb=30 235 525 540,clip,scale=0.45]{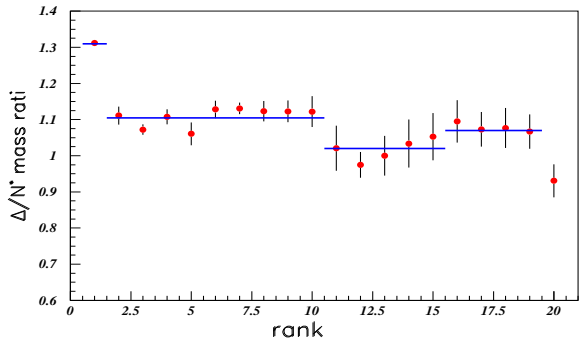}}
\caption{Color on line. Ratio of baryonic PDG $\Delta/N^{*}$ masses.}
\end{center}
\end{figure}
\begin{figure}[h]
\begin{center}
\hspace*{-3.mm}
\vspace*{1.mm}
\scalebox{1}[0.9]{
\includegraphics[bb=45 240 530 540,clip,scale=0.45]{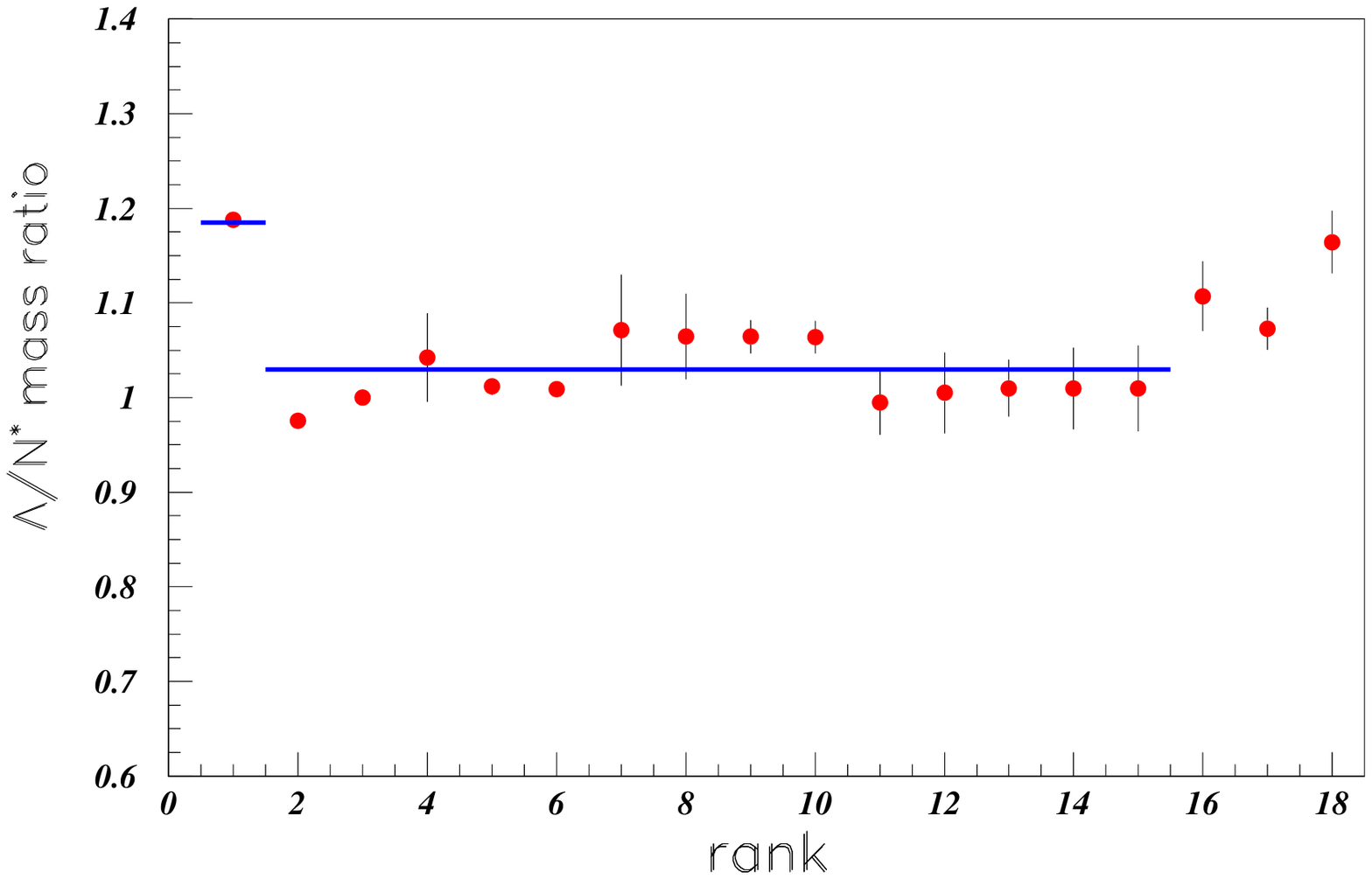}}
\vspace*{-2.mm}
\caption{Color on line. Ratio of baryonic PDG $\Lambda/N^{*}$ masses.}
\end{center}
\end{figure}
\begin{figure}[h]
\begin{center}
\hspace*{-3.mm}
\vspace*{1.mm}
\scalebox{1}[0.9]{
\includegraphics[bb=45 240 530 540,clip,scale=0.45]{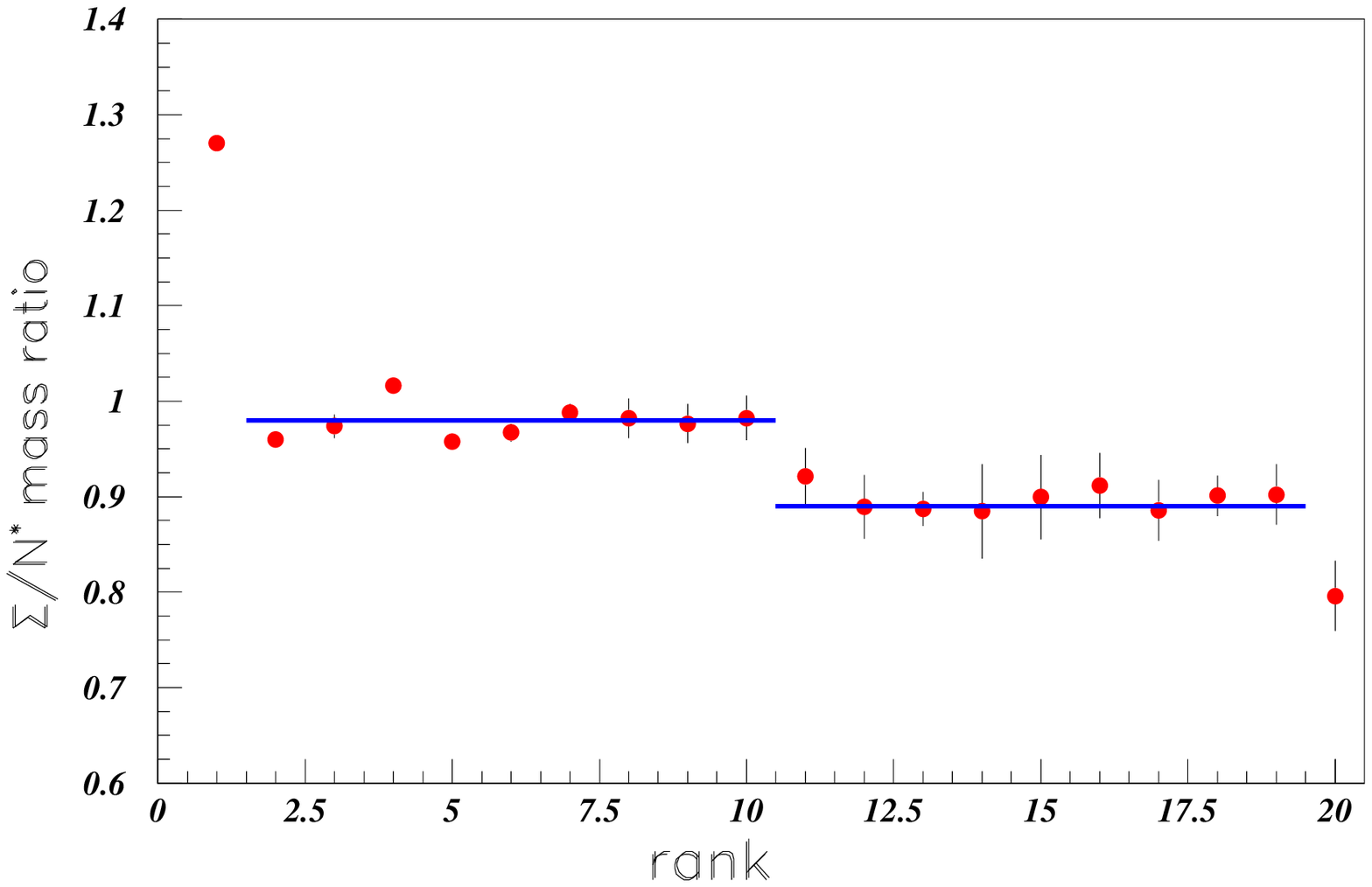}}
\vspace*{-2.mm}
\caption{Color on line. Ratio of baryonic PDG $\Sigma/N^{*}$ masses.}
\end{center}
\end{figure}
\begin{figure}[h]
\begin{center}
\hspace*{-3.mm}
\vspace*{1.mm}
\scalebox{1}[0.9]{
\includegraphics[bb=45 240 530 540,clip,scale=0.45]{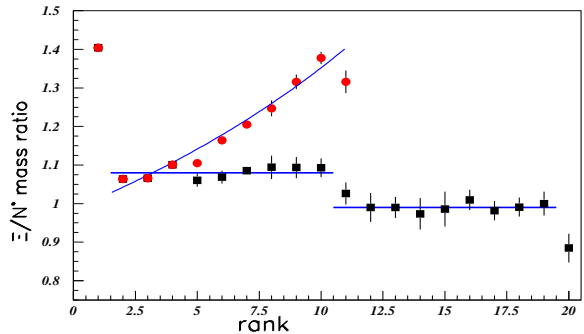}}
\vspace*{-2mm}
\caption{Color on line.  The red circles show the ratio of baryonic PDG $\Xi/N^{*}$ masses. Black squares show the same ratio after a proposed introduction of some arbitrary missing $\Xi$ baryonic masses (see text).}
\end{center}
\end{figure}
\begin{figure}[h]
\begin{center}
\hspace*{-3.mm}
\vspace*{1.mm}
\scalebox{1}[0.9]{
\includegraphics[bb=45 240 530 540,clip,scale=0.45]{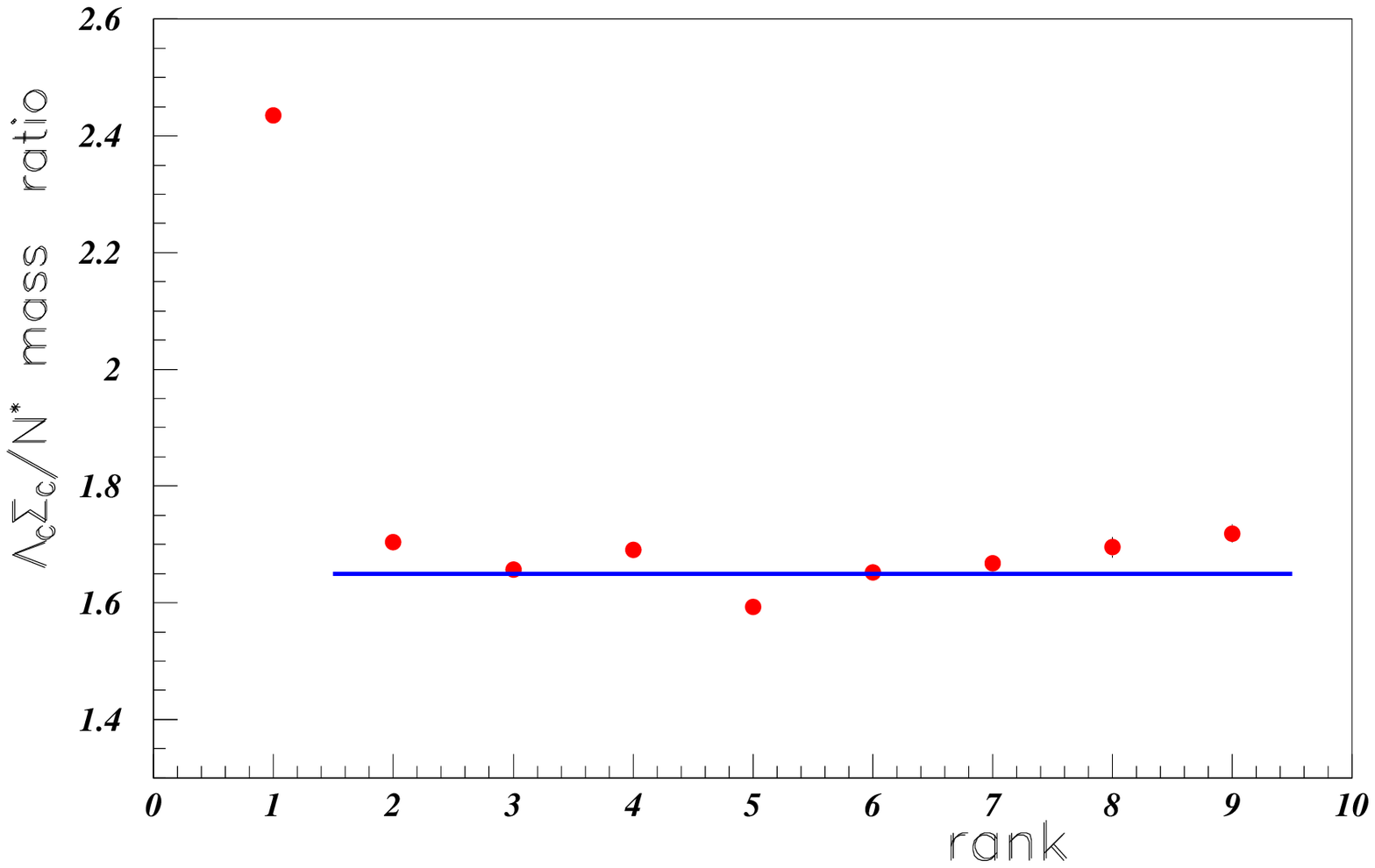}}
\vspace*{-2mm}
\caption{Color on line. Ratio of baryonic PDG $\Lambda_{C}\Sigma_{C}/N^{*}$ masses.}
\end{center}
\end{figure}
\begin{figure}[b]
\begin{center}
\hspace*{-3.mm}
\vspace*{1.mm}
\scalebox{1}[0.9]{
\includegraphics[bb=38 240 515 540,clip,scale=0.45]{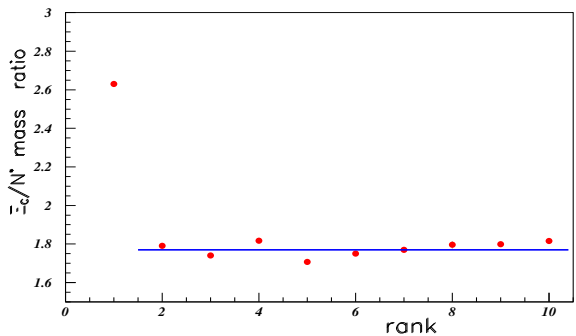}}
\vspace*{-2.mm}
\caption{Color on line. Ratio of baryonic PDG $\Xi_{C}/N^{*}$ masses.}
\end{center}
\end{figure}
\subsection{Baryonic masses compared to $\Delta$ masses}
Figure~8 shows the ratio of PDG $\Sigma/\Delta$ masses, $\Lambda^{0}/\Delta$ masses, and $\Lambda^{+}_{C}\Sigma^{++}_{C}/\Delta$ masses (full red circles).
\begin{figure}[h]
\begin{center}
\hspace*{-3.mm}
\vspace*{1.mm}
\scalebox{1}[0.9]{
\includegraphics[bb=16 94  521 546,clip,scale=0.45]{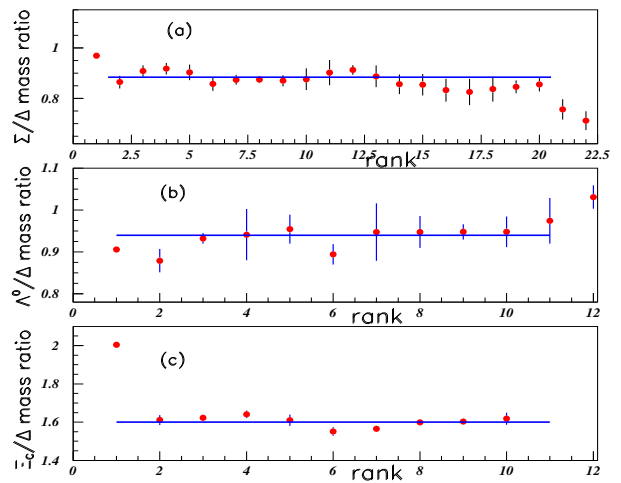}}
\vspace*{-2.mm}
\caption{Color on line. Inserts (a), (b), and (c), show respectively the ratio of baryonic PDG $\Sigma/\Delta$ masses, $\Lambda^{0}/\Delta$ masses, and $\Xi_{C}/\Delta$ masses.}
\end{center}
\end{figure}
Figure~9 shows the ratio of $\Xi/\Delta$ masses. The same comments apply, as those concerning the $\Xi/N^{*}$ mass ratio (figure~5). Figure 10 and figure 11 show the ratio of PDG $\Lambda_{C}\Sigma_{C}/\Delta$baryonic masses. We observe in both cases, a nice horizontal line, except for the yrast mass ratio.
\begin{figure}[h]
\begin{center}
\hspace*{-3.mm}
\vspace*{1.mm}
\scalebox{1}[0.9]{
\includegraphics[bb=31 241 516 545,clip,scale=0.45]{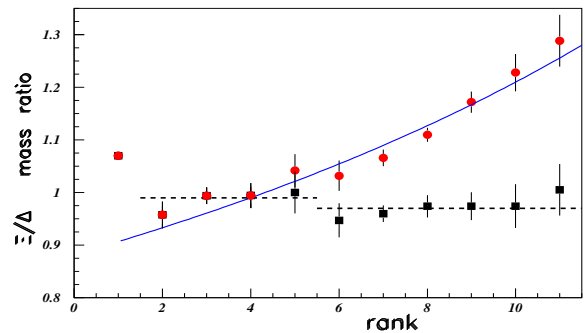}}
\vspace*{-2.mm}
\caption{Color on line. Red circles show the ratio of baryonic PDG $\Xi/\Delta$ masses. Black squares show the same ratio after the proposed introduction of arbitrary missing $\Xi$ baryonic masses (see text).}
\end{center}
\end{figure}
\begin{figure}[h]
\begin{center}
\hspace*{-3.mm}
\vspace*{1.mm}
\scalebox{1}[0.9]{
\includegraphics[bb=38 240 530 540,clip,scale=0.45]{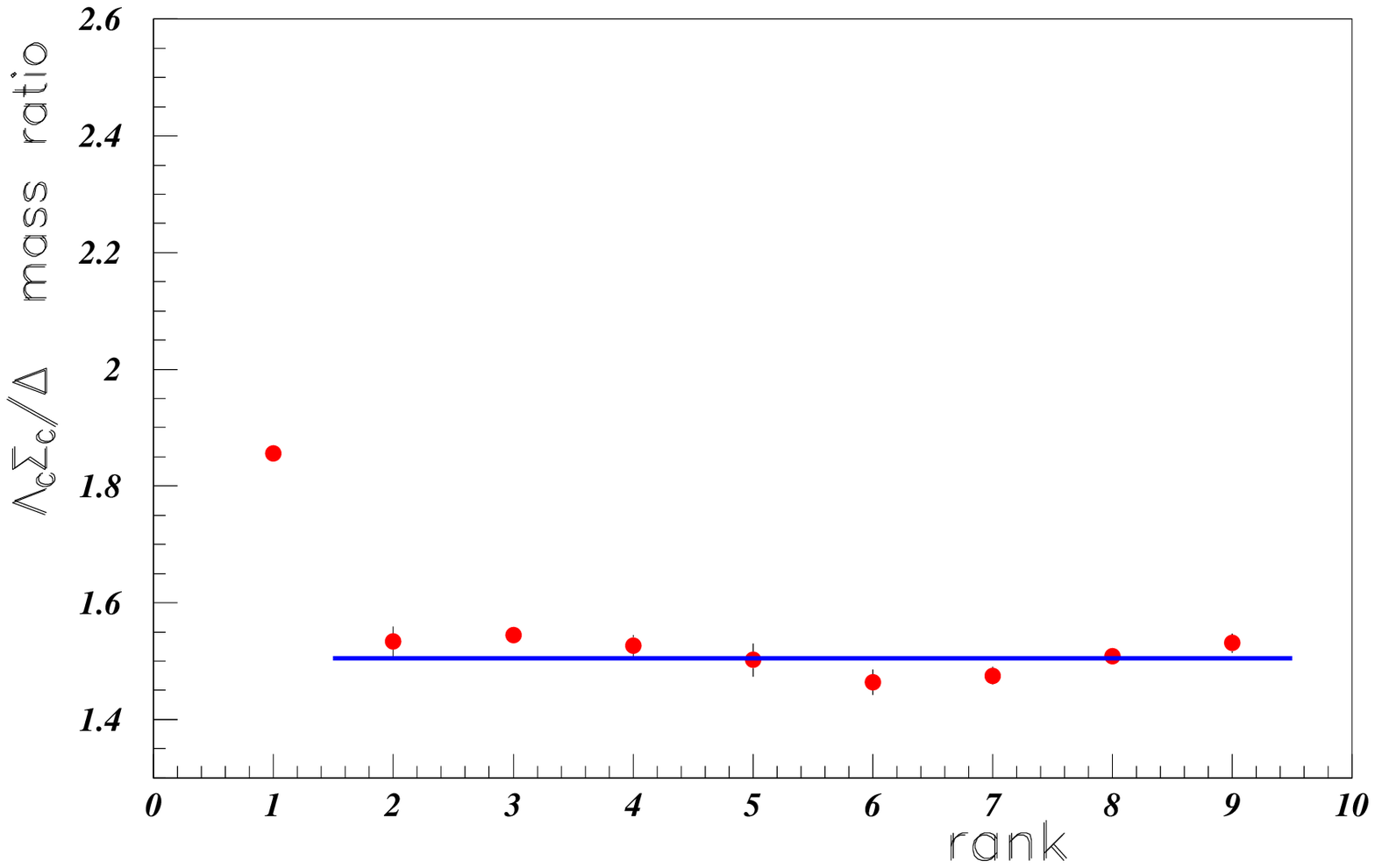}}
\vspace*{-2.mm}
\caption{Color on line. Red circles show the ratio of baryonic PDG 
$\Lambda_{C}\Sigma_{C}/\Delta$ masses.}
\end{center}
\end{figure}
\begin{figure}[h]
\begin{center}
\hspace*{-3.mm}
\vspace*{1.mm}
\scalebox{1}[0.9]{
\includegraphics[bb=38 240 530 540,clip,scale=0.45]{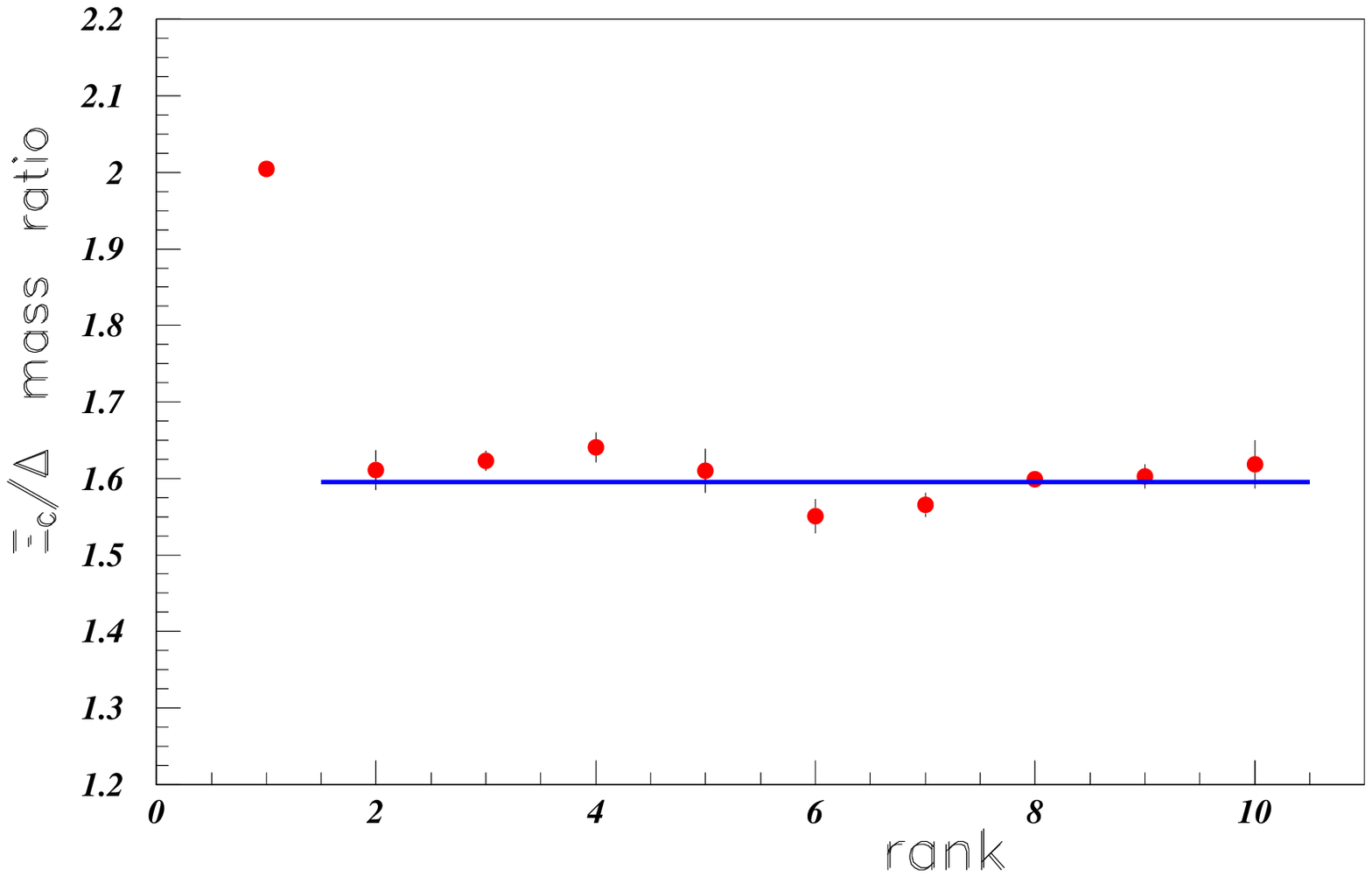}}
\vspace*{-2.mm}
\caption{Color on line. Red circles show the ratio of baryonic PDG  $\Xi_{C}/\Delta$ masses.}
\end{center}
\end{figure}
\subsection{Baryonic masses compared to $\Lambda$ masses}
Figure~12 shows the ratio of $\Sigma/\Lambda$ baryonic masses. In both species, more data than before exists, allowing to draw the figure up to rank 18. We observe several horizontal lines decreasing for increasing rank by a rather constant gap.
\begin{figure}[h]
\begin{center}
\hspace*{-3.mm}
\vspace*{1.mm}
\scalebox{1}[0.9]{
\includegraphics[bb=38 240 530 540,clip,scale=0.45]{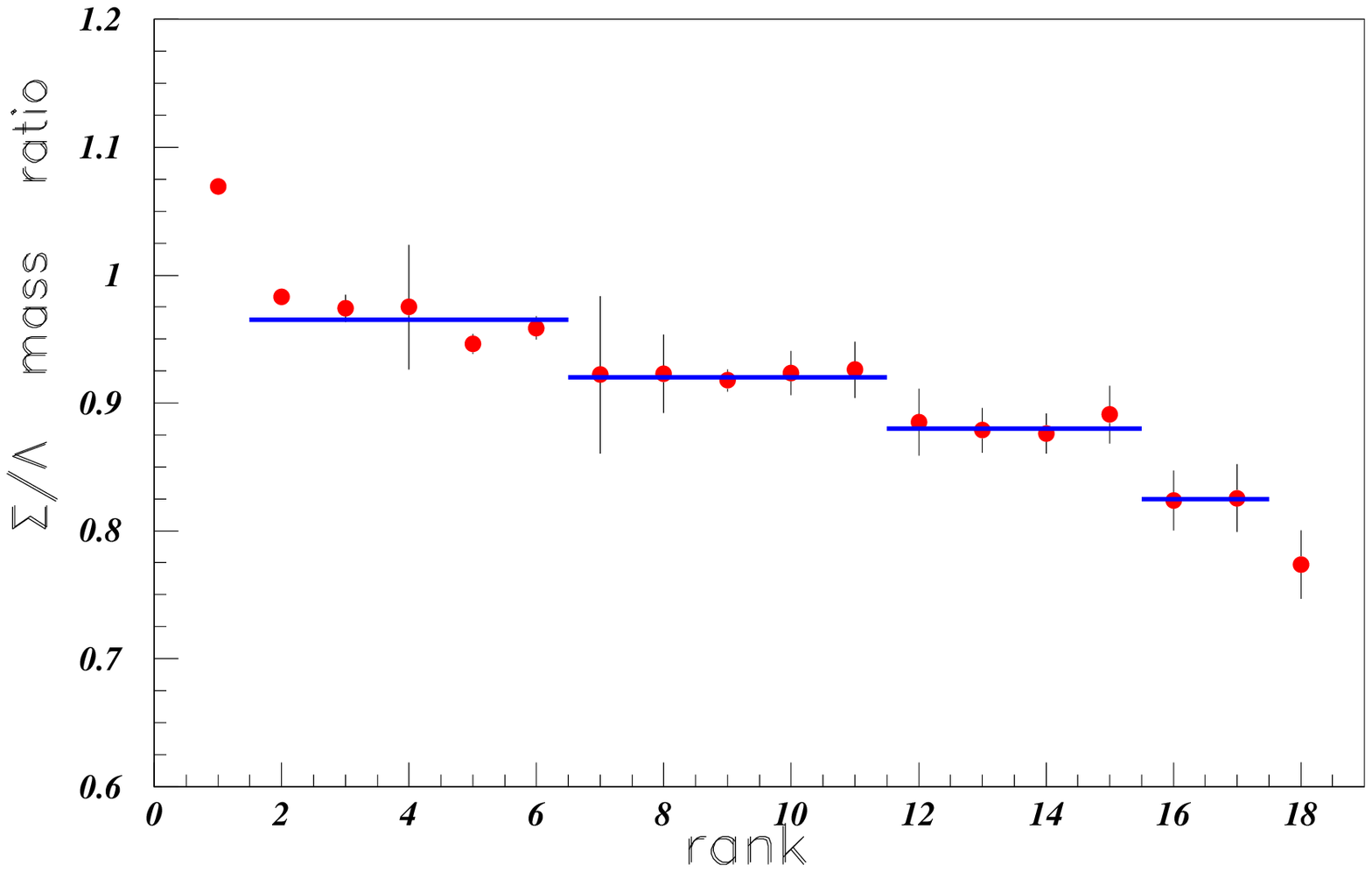}}
\vspace*{-2.mm}
\caption{Color on line. Red circles show the ratio of baryonic PDG  $\Sigma/\Lambda$ masses.}
\end{center}
\end{figure}
Figure~13 shows the ratio of $\Xi/\Lambda$ masses. Here again, the comments of figure~5 apply.
\begin{figure}[h]
\begin{center}
\hspace*{-3.mm}
\vspace*{1.mm}
\scalebox{1}[0.9]{
\includegraphics[bb=38 240 530 540,clip,scale=0.45]{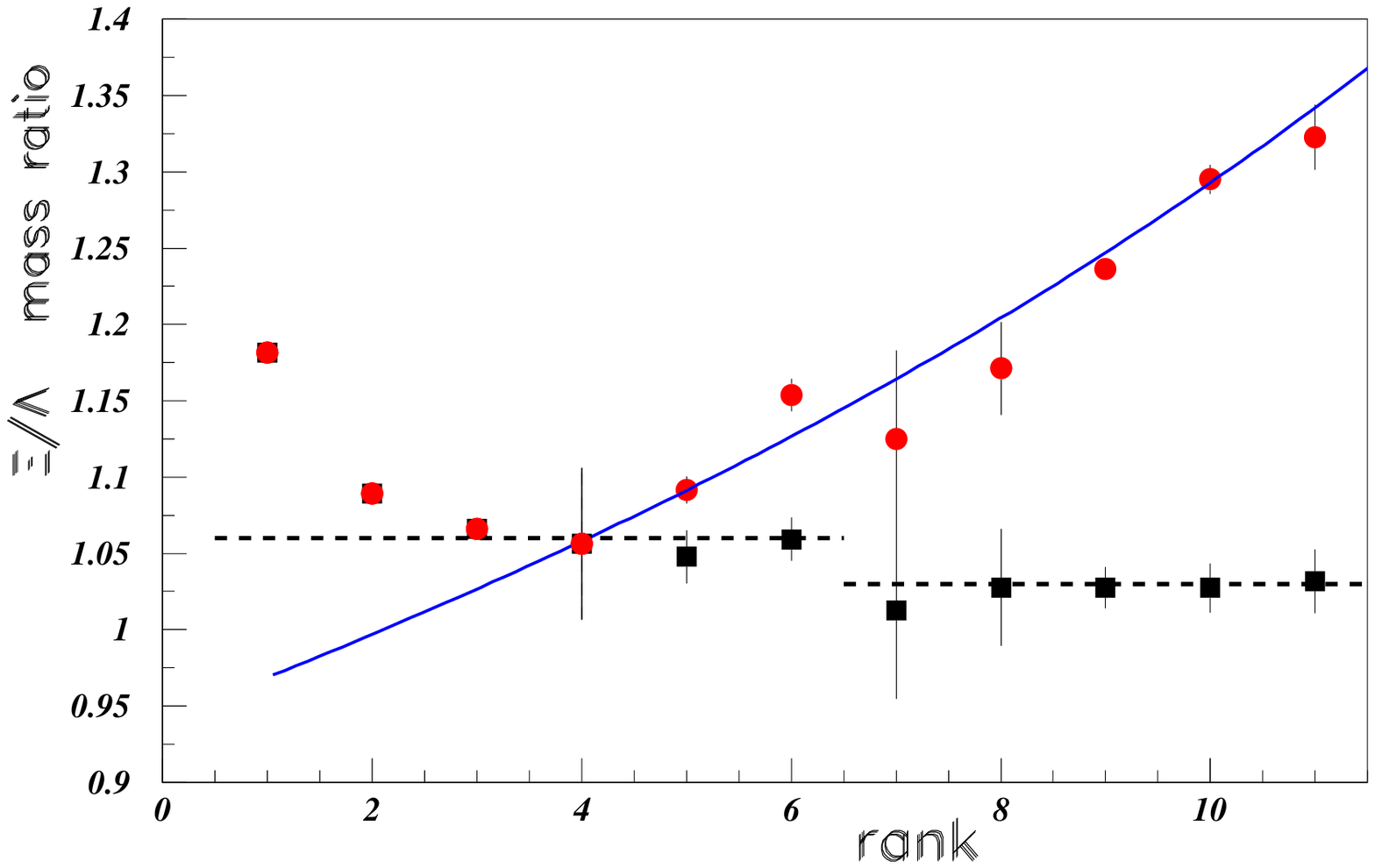}}
\vspace*{-2.mm}
\caption{Color on line. Red circles show the ratio of baryonic PDG  $\Xi/\Lambda$ masses. The black squares show the same ratio after introduction of proposed missing $\Xi$ baryonic masses (see text).}
\end{center}
\end{figure}
Figures 14 and 15 show the ratios of $\Lambda^{+}_{C}\Sigma^{++}_{C}$ and $\Xi_{C}$ over $\Lambda$ baryons. 
\begin{figure}[h]
\begin{center}
\hspace*{-3.mm}
\vspace*{1.mm}
\scalebox{1}[0.9]{
\includegraphics[bb=38 240 530 540,clip,scale=0.45]{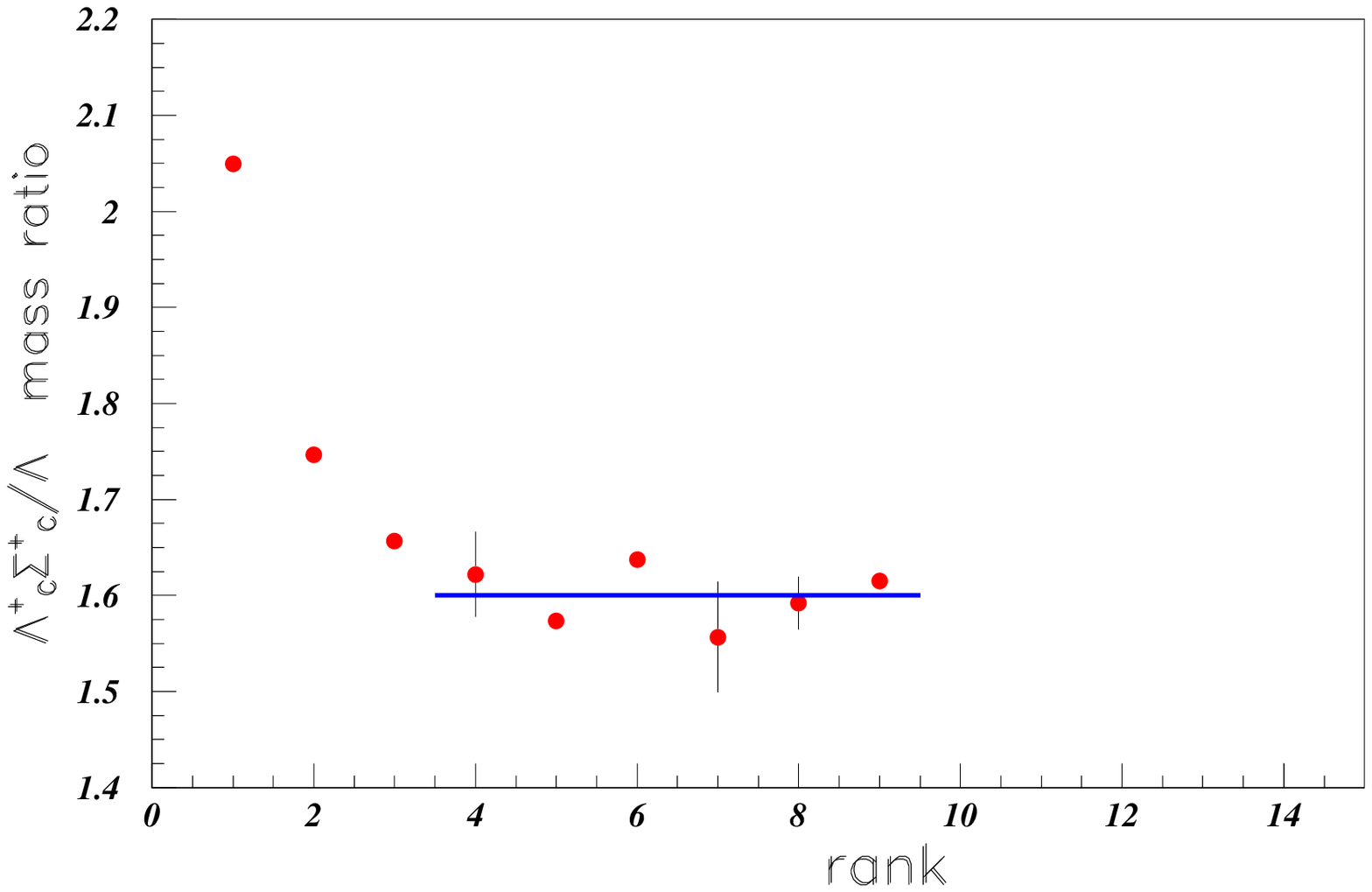}}
\vspace*{-2.mm}
\caption{Color on line. Red circles show the ratio of baryonic PDG  
$\Lambda^{+}_{C}\Sigma^{+}_{C}/\Lambda$ masses.}
\end{center}
\end{figure}
\begin{figure}[h]
\begin{center}
\hspace*{-3.mm}
\vspace*{1.mm}
\scalebox{1}[0.9]{
\includegraphics[bb=38 240 530 540,clip,scale=0.45]{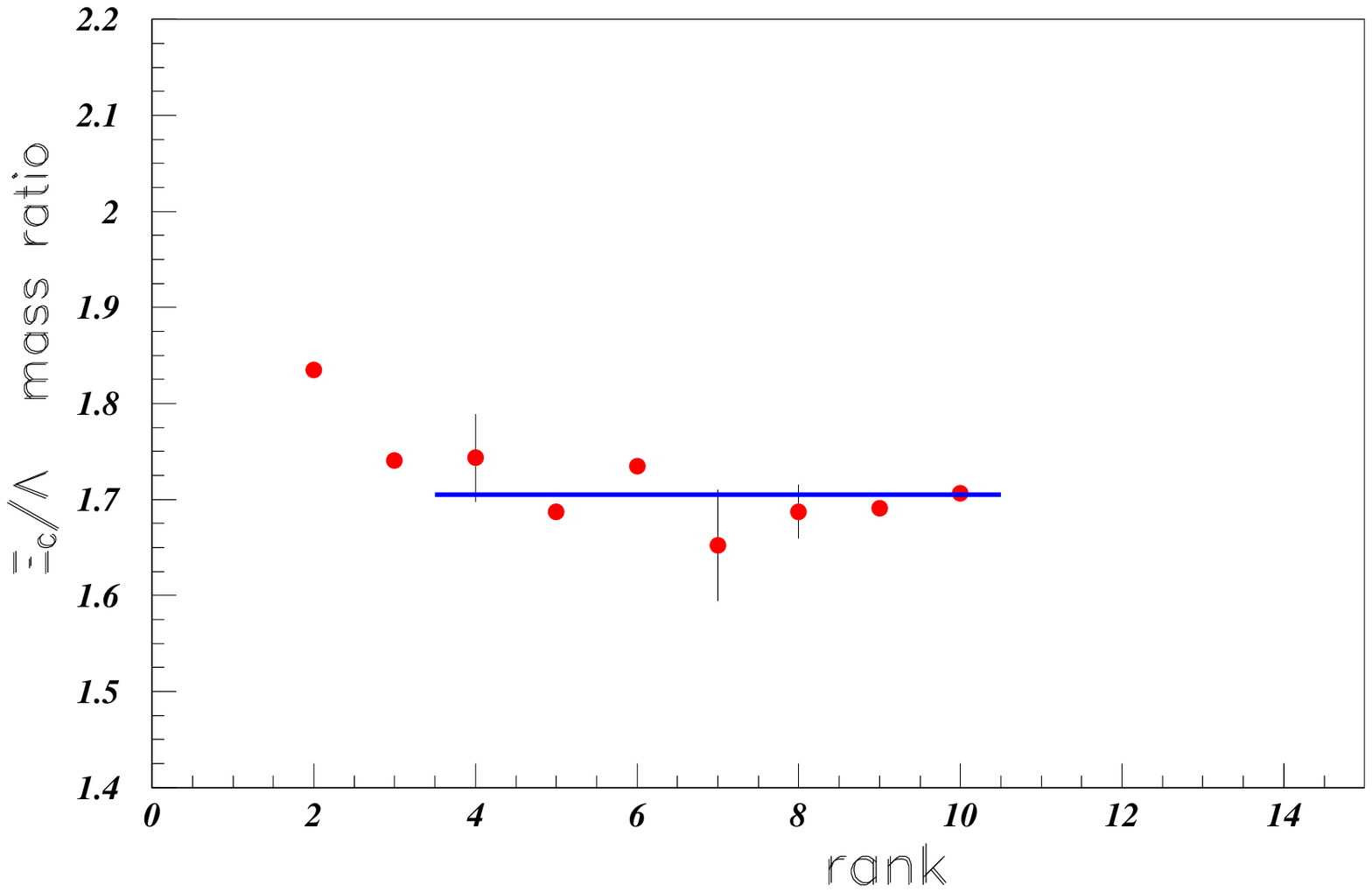}}
\vspace*{-2.mm}
\caption{Color on line. Red circles show the ratio of baryonic PDG  $\Xi_{C}/\Lambda$ masses.}
\end{center}
\end{figure}
\subsection{Baryonic masses compared to $\Sigma$ masses}
Figure~16 shows the ratio of $\Xi/\Sigma$ baryonic masses. The same comments, as those concerning the dicussion of figure~5, apply here, since  again the ratio involves the $\Xi$ baryons. Figures~17 and 18 show the ratio of $\Lambda^{+}_{C}\Sigma^{++}_{C} / \Sigma$ and $\Xi_{C} / \Sigma$ baryonic masses.
\begin{figure}[h]
\begin{center}
\hspace*{-3.mm}
\vspace*{1.mm}
\scalebox{1}[0.9]{
\includegraphics[bb=38 240 530 540,clip,scale=0.45]{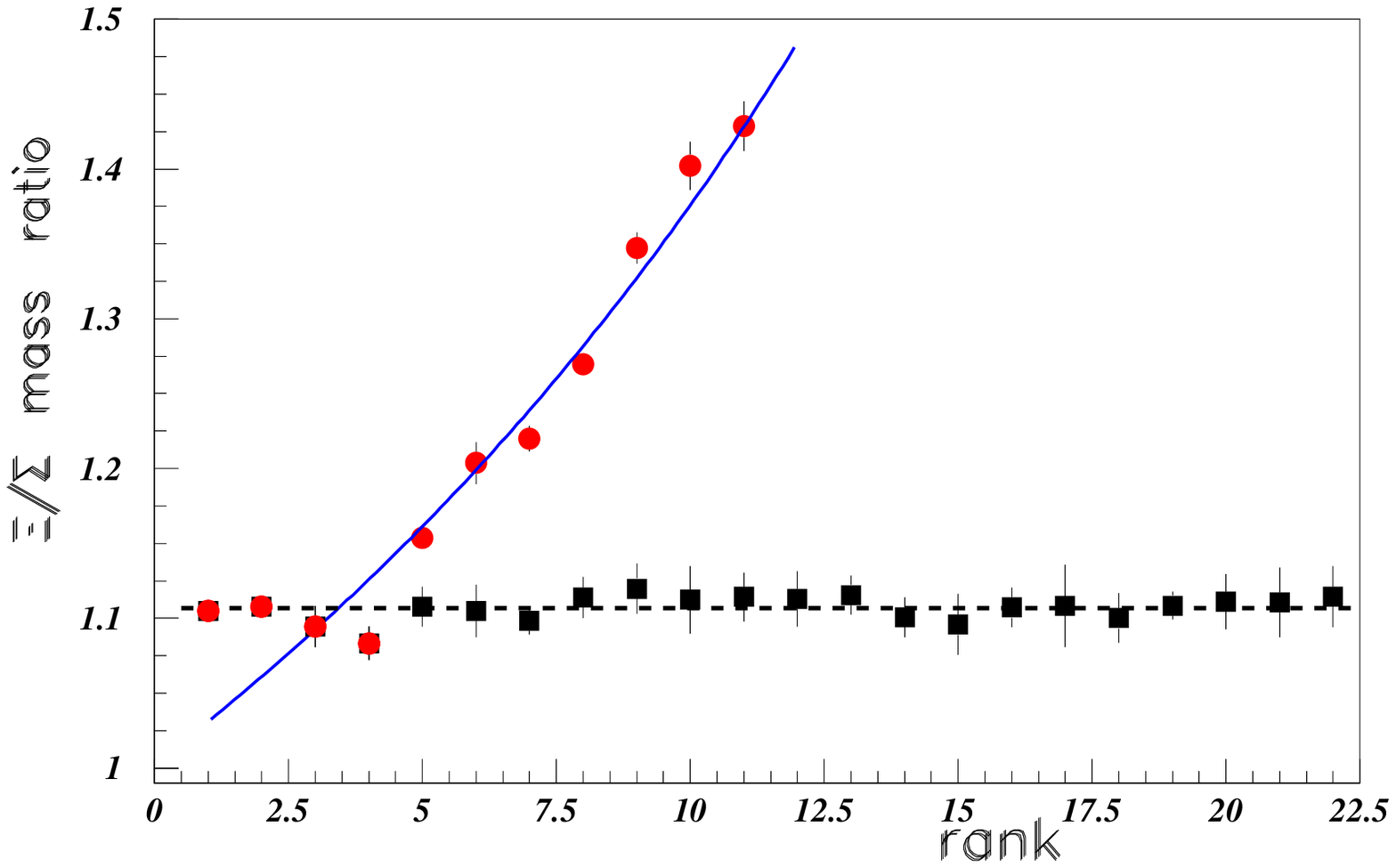}}
\vspace*{-2.mm}
\caption{Color on line. Red circles show the ratio of baryonic PDG  $\Xi/\Sigma$ masses.
Black squares show the same ratio after proposed introduction of new $\Xi$ baryonic masses as in figure~13 (see text).}
\end{center}
\end{figure}
\begin{figure}[h]
\begin{center}
\hspace*{-3.mm}
\vspace*{1.mm}
\scalebox{1}[0.9]{
\includegraphics[bb=38 240 530 540,clip,scale=0.45]{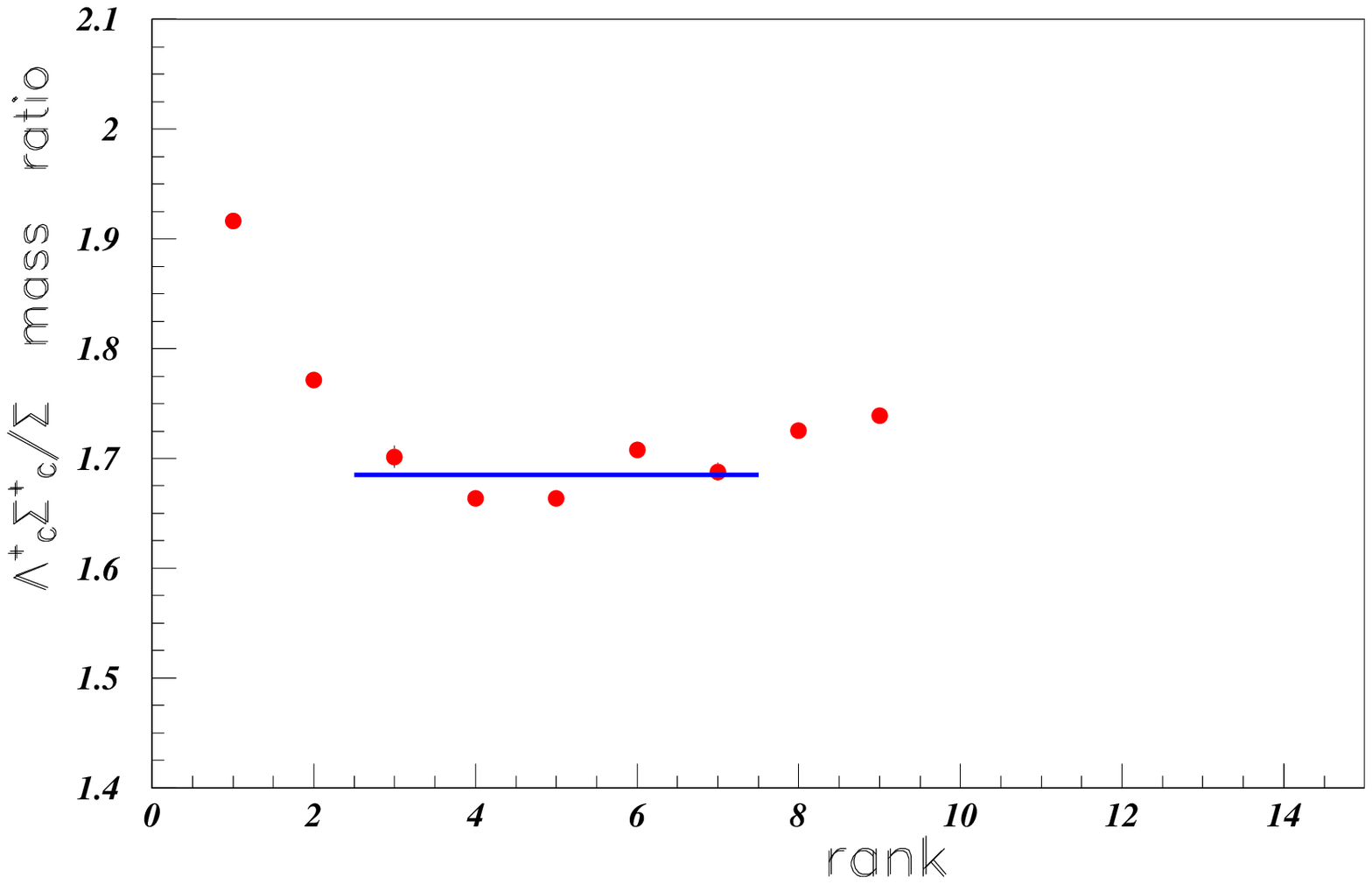}}
\vspace*{-2.mm}
\caption{Color on line. Red circles show the ratio of baryonic PDG  $\Lambda^{+}_{C}\Sigma^{+}_{C}/\Sigma$ masses.}
\end{center}
\end{figure}
\begin{figure}[h]
\begin{center}
\hspace*{-3.mm}
\vspace*{1.mm}
\scalebox{1}[0.9]{
\includegraphics[bb=38 240 530 540,clip,scale=0.45]{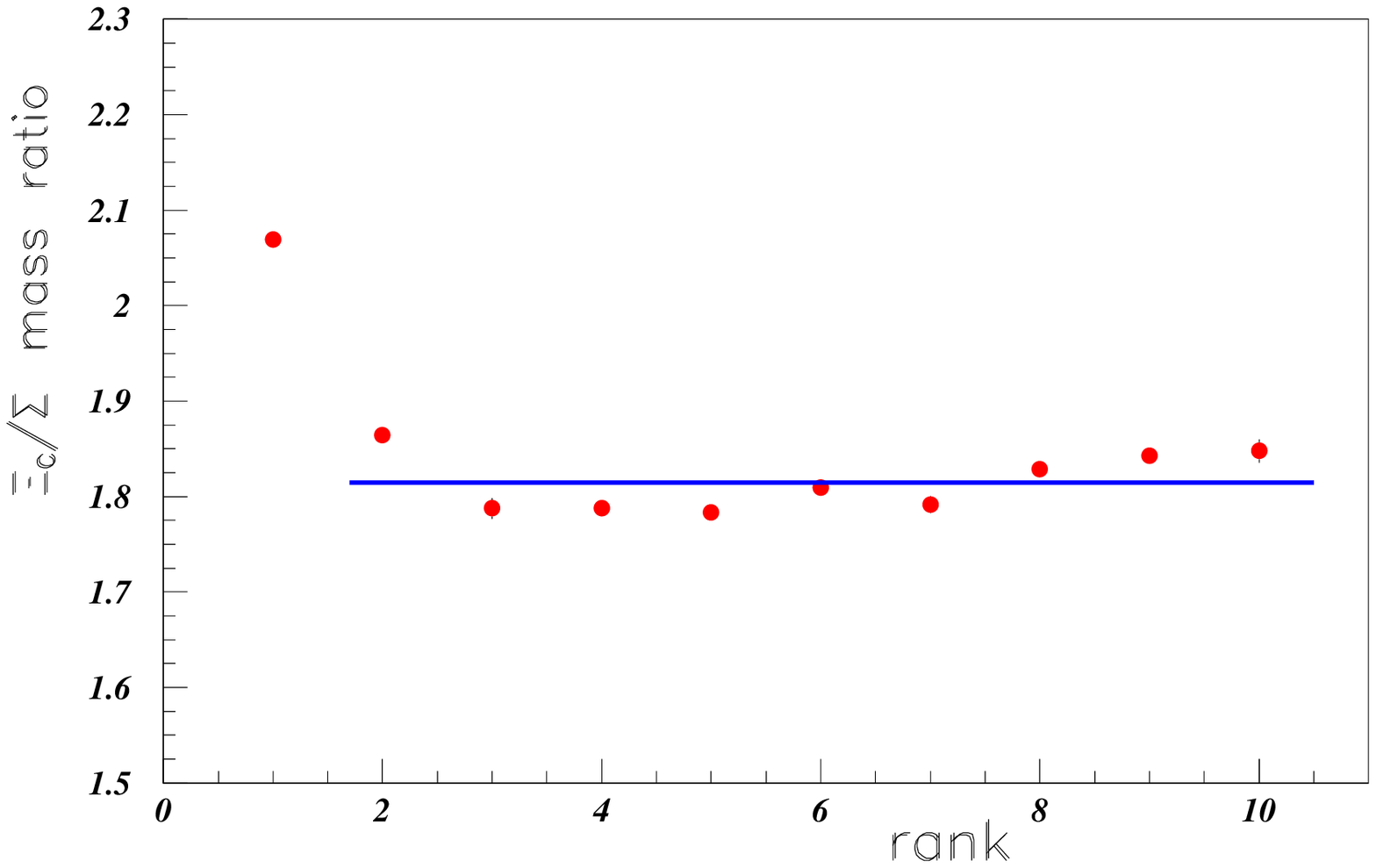}}
\vspace*{-2.mm}
\caption{Color on line. Red circles show the ratio of baryonic PDG  $\Xi_{C}/\Sigma$ masses.}
\end{center}
\end{figure}
\subsection{Baryonic masses compared to $\Xi$ masses}
Figures~19 and 20 show the ratios of $\Lambda^{+}_{C}\Sigma^{++}_{C}$ and $\Xi_{C}$ over $\Xi$ baryonic masses. The same comments, as those concerning the dicussion of figure~5, apply here, since  again both ratios involve the $\Xi$ baryons. The lack of many masses in the $\Xi$ baryon species, explains why the distributions decrease.
\begin{figure}[h]
\begin{center}
\hspace*{-3.mm}
\vspace*{1.mm}
\scalebox{1}[0.9]{
\includegraphics[bb=38 240 530 540,clip,scale=0.45]{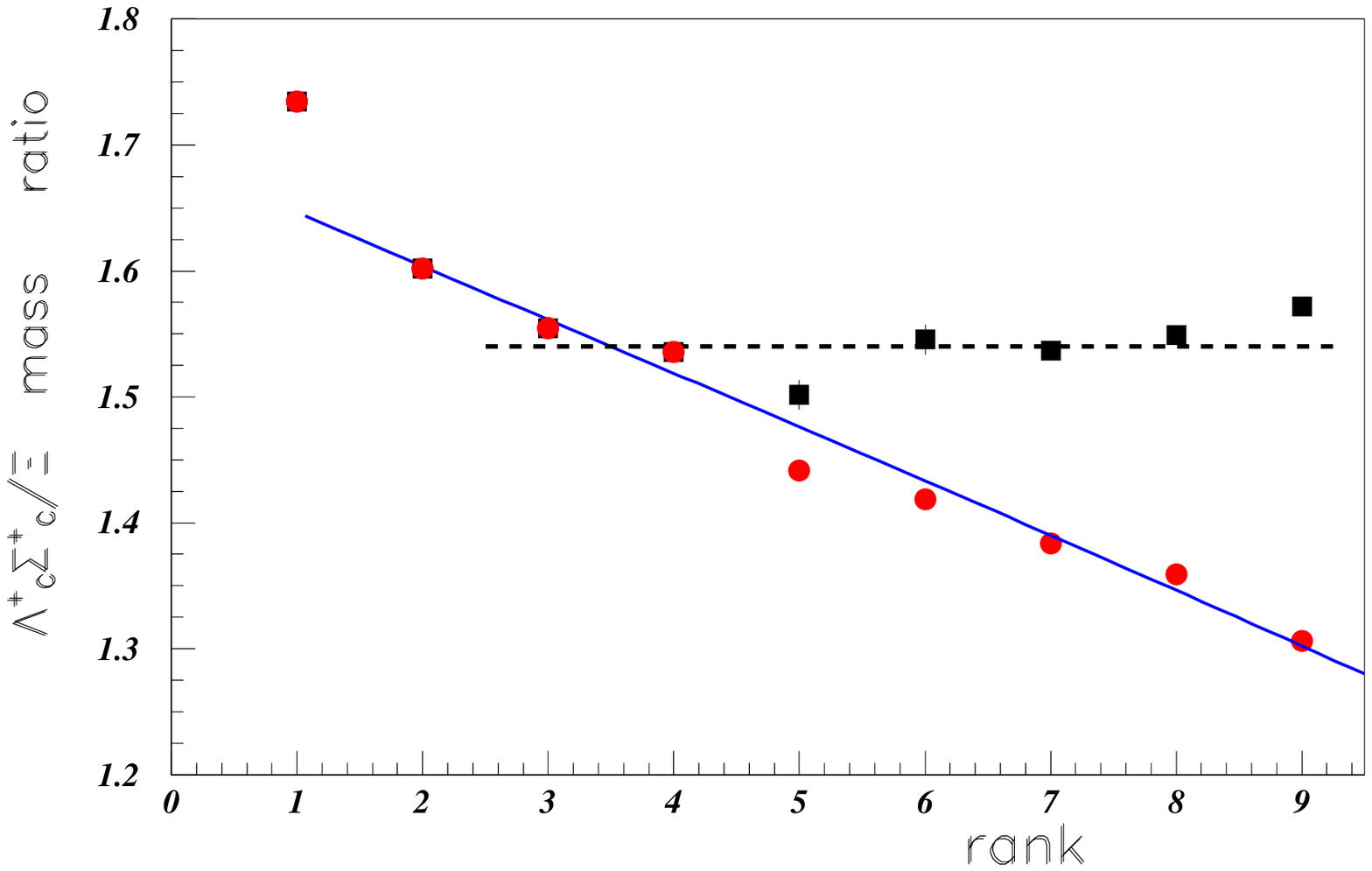}}
\vspace*{-2.mm}
\caption{Color on line. Red circles show the ratio of baryonic PDG 
 $\Lambda^{+}_{C}\Sigma^{+}_{C}/\Xi$ masses. Black squares show the same ratio after proposed introduction of new $\Xi$ baryonic masses as in figure~13 (see text).}
\end{center}
\end{figure}
\begin{figure}[h]
\begin{center}
\hspace*{-3.mm}
\vspace*{1.mm}
\scalebox{1}[0.9]{
\includegraphics[bb=38 240 530 540,clip,scale=0.45]{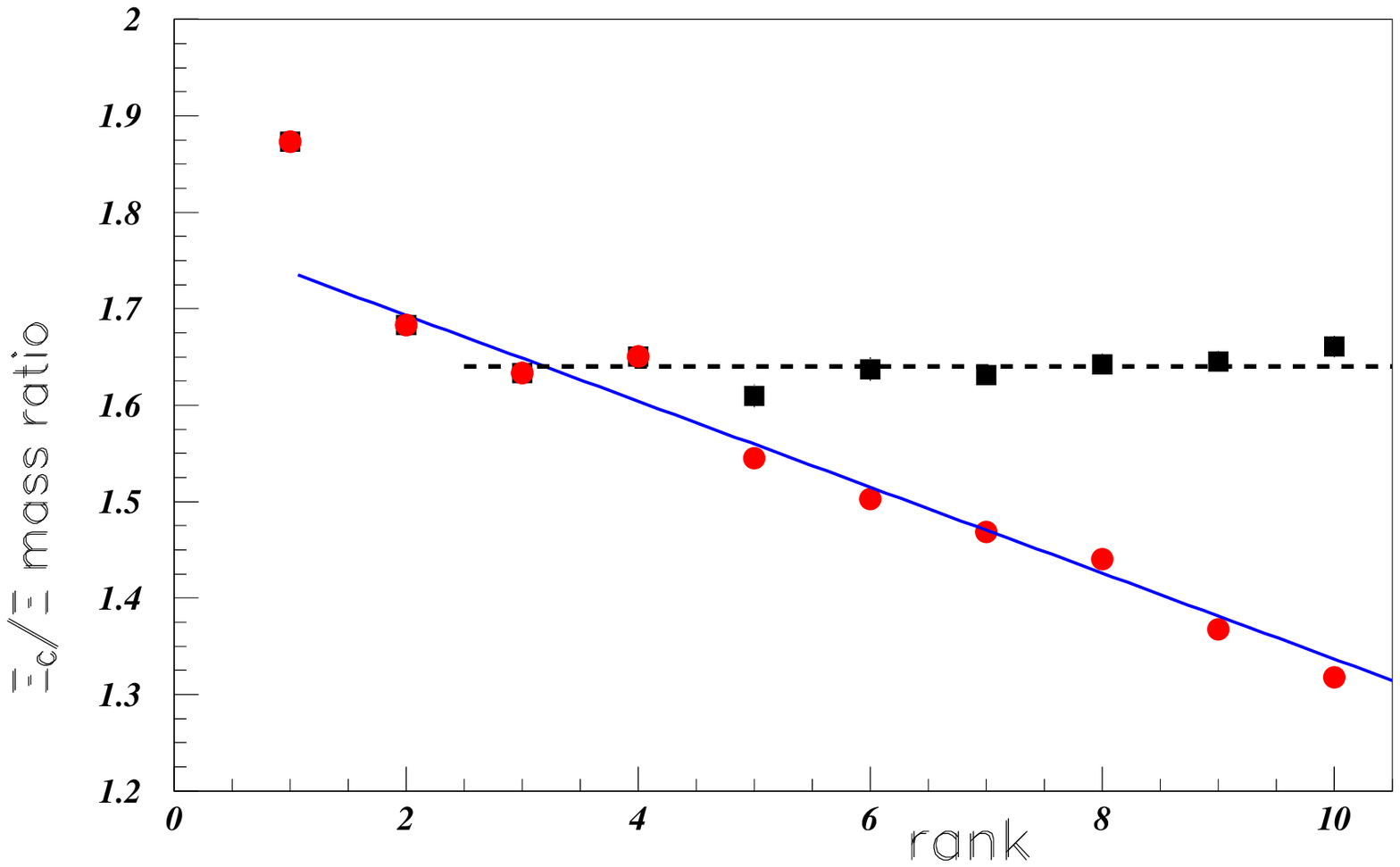}}
\vspace*{-2.mm}
\caption{Color on line. Red circles show the ratio of baryonic PDG 
 $\Xi_{C}/\Xi$ masses. Black squares show the same ratio after proposed introduction of new $\Xi$ baryonic masses as in figure~13 (see text).}
\end{center}
\end{figure}
\subsection{Baryonic $\Xi_{C}$ masses compared to $\Lambda_{C}$ masses}
Figure 21 shows the ratio of $\Xi_{C}$ masses compared to $\Lambda_{C}\Sigma^{++}_{C}$ masses. These data display an horizontal line, indicating with a rather great probability, that there is no one missing mass in both species.
\begin{figure}[h]
\begin{center}
\hspace*{-3.mm}
\vspace*{1.mm}
\scalebox{1}[0.9]{
\includegraphics[bb=38 240 530 540,clip,scale=0.45]{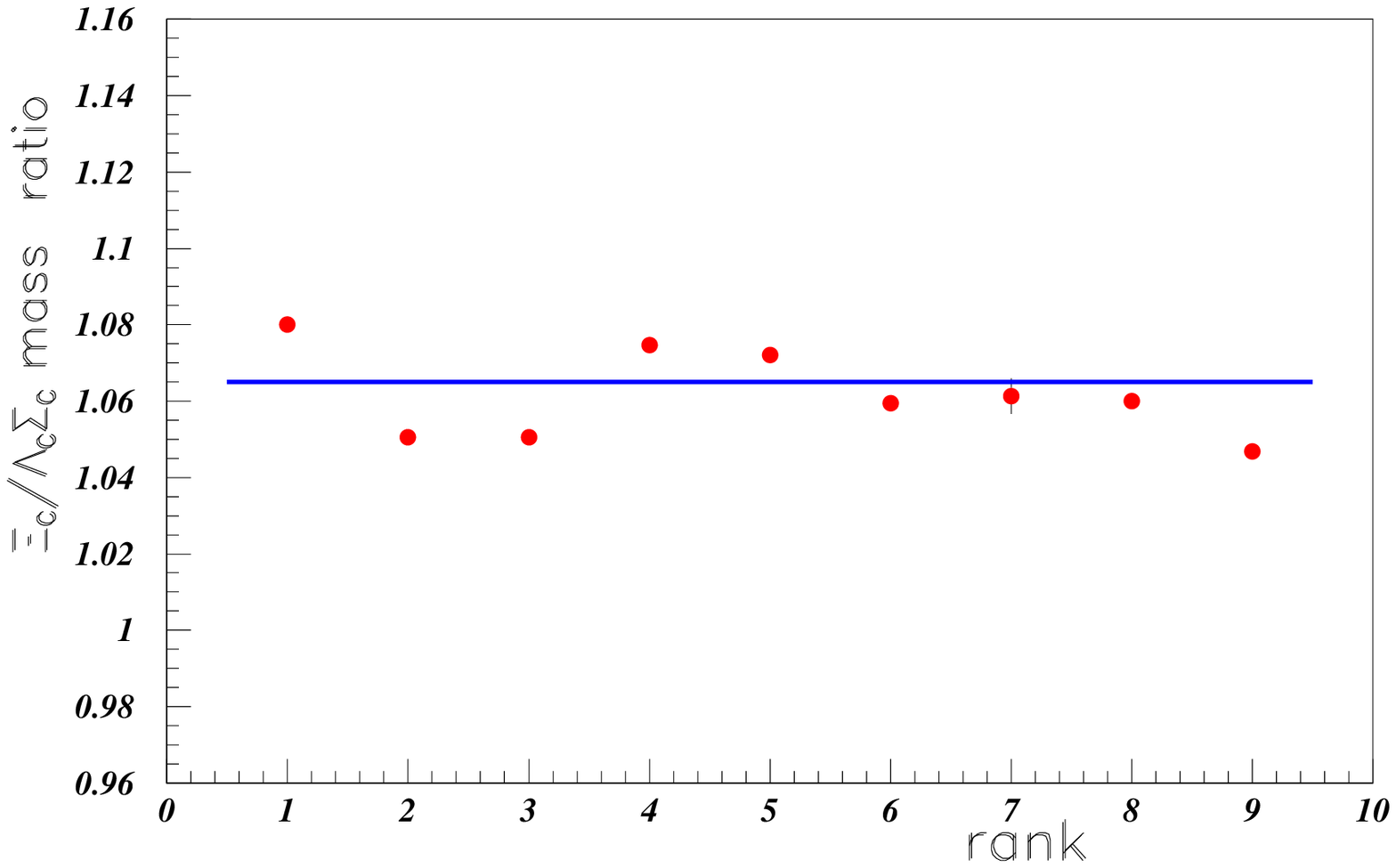}}
\vspace*{-2.mm}
\caption{Color on line. Red circles show the ratio of baryonic PDG 
 $\Xi_{C}/\Lambda_{C}\Sigma_{C}$ masses.}
\end{center}
\end{figure}
\section{Discussion}
A proportionality is observed between excited particle masses of all baryonic species, at least up to rank 10, except for the $\Xi$ baryons. We have therefore supposed that this is a general property, and naturally we associate the differences observed in all ratios involving the $\Xi$ baryons,  with a lack of some $\Xi$ baryonic masses. 
\subsection{Discussion concerning the $\Xi$ baryon masses}
 The same set of tentatively introduced $\Xi$ baryonic masses, allows to obtain proportionality with all other baryonic species. We observe that this is the case in all figures involving $\Xi$ baryons. In addition to the PDG $\Xi$ masses:
M =  1318.28~MeV, 1531.8~MeV, 1620~MeV, 1690~MeV, 1823~MeV, 1950~MeV, 2025~MeV, 2120~MeV, 2250~MeV, 2370~MeV, and 2500~MeV, the proportionality is obtained after the arbitrary introduction of the following masses:  1750~MeV, 1790~MeV, 1860~MeV, 1870~MeV, 1880~MeV, 1970~MeV, 1980~MeV, 2060~MeV, 2150~MeV, 2200~MeV, 2300~MeV, 2310~MeV, 2340~MeV, and 2360~MeV. 

It is of course out not possible, to attribute definitively these 14 supplementary introduced masses, to exact missing $\Xi$ baryonic masses. They must be used, rather, as an indication of the number and approximate masses,
 that have to be observed by specific experiments still to be done.

The $\Xi$ baryon masses are more precise, around M~$\approx$~2~GeV, than the N$^{*}$ masses. Taking into account the small shifts between adjacent introduced additionnal $\Xi$ masses, we attribute for them an unprecision $\Delta$(M)~=~$\pm$~20~MeV.

The result of this increase of $\Xi$ mass data, is shown in figures 5, 9, 13, 16, 19, and 20, by black squares. The same set of new proposed $\Xi$ masses allows to notably improve all ratios between different baryonic species. These masses are introduced in red in table~1.
\subsection{Discussion concerning the $N^{*}$ baryon masses}
In the same way, as done for $\Xi$ baryons, a careful look on the figures where the baryon masses are compared to $N^{*}$ masses, indicates a possible improvement after a tentative introduction of several masses in the range  1720$\le~M~\le$~1900~MeV.
For different species where  at least 11 masses are known, the comparison to the $N^{*}$ masses, allows us to observe a shift starting always at rank 11. 
Such shift may indicate a lack of one (or a few) $N^{*}$ masses in the range 1720$\le$M$\le$1900~MeV, where there is no $N^{*}$ baryonic mass in the PDG table. Indeed in the lower range 
1650$\le$M$\le$1720~MeV, there are six masses reported in the same table.
The arbitrary introduction of the following masses: M$\approx$ 1750~MeV, 1780~MeV, and 1820~MeV allows us to improve the linearity between the various baryonic species masses. 

The precision on N$^{*}$ masses is often poor for masses $\ge$~1700~MeV. When not given in PDG, we take into account the various experimental masses to attribute a mass unprecision. The precision on all tentatively introduced N$^{*}$ masses, is arbitrarily put to $\Delta$~M~=~$\pm$50~MeV. 

Figure~22, shows the new $\Lambda / N^{*}$ mass ratio, which has to be compared to figure~3. Figure~25 shows the $\Delta/N^{*}$ mass ratio, after the introduction of additionnal N$^{*}$ masses.
\subsection{Discussion concerning the $\Lambda$ baryonic masses}
Another observation of previous figures shows that there is missing mass(es) in the $\Lambda$ mass table in the range 1690$\le$M~$\le$1800~MeV. Indeed this appear in figure~3
($\Lambda/N^{*}$ mass ratio), figure~13 ($\Xi/\Lambda$ mass ratio), and figure~12
($\Sigma/\Lambda$ mass ratio).  The tentatively introduction of three additionnal $\Lambda$ masses: M = 1720~MeV, 1750~MeV, and 1780~MeV allows to improve the proportionality of baryonic mass ratios up to rank = 18 as shown in the figures 26, 27, and 28.
\begin{figure}[h]
\begin{center}
\hspace*{-3.mm}
\vspace*{1.mm}
\scalebox{1}[0.9]{
\includegraphics[bb=38 240 530 540,clip,scale=0.45]{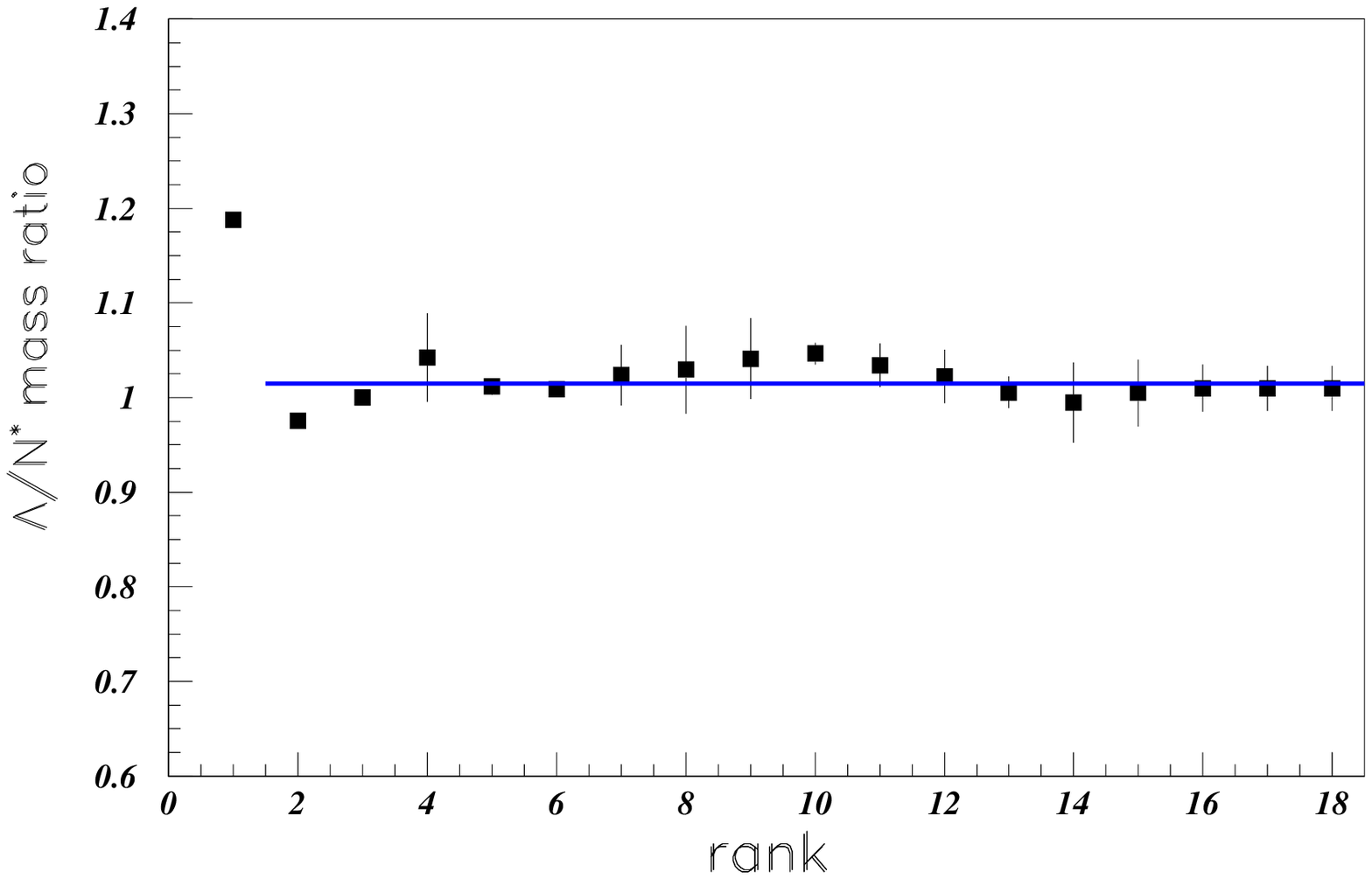}}
\vspace*{-2.mm}
\caption{Black squares show the ratio of baryonic  $\Lambda/N^{*}$  masses. Here, both PDG $N^{*}$ and $\Lambda$ data are increased by additionnal masses (see text).}
\end{center}
\end{figure}
\begin{figure}[h]
\begin{center}
\hspace*{-3.mm}
\vspace*{1.mm}
\scalebox{1}[0.9]{
\includegraphics[bb=38 240 530 540,clip,scale=0.45]{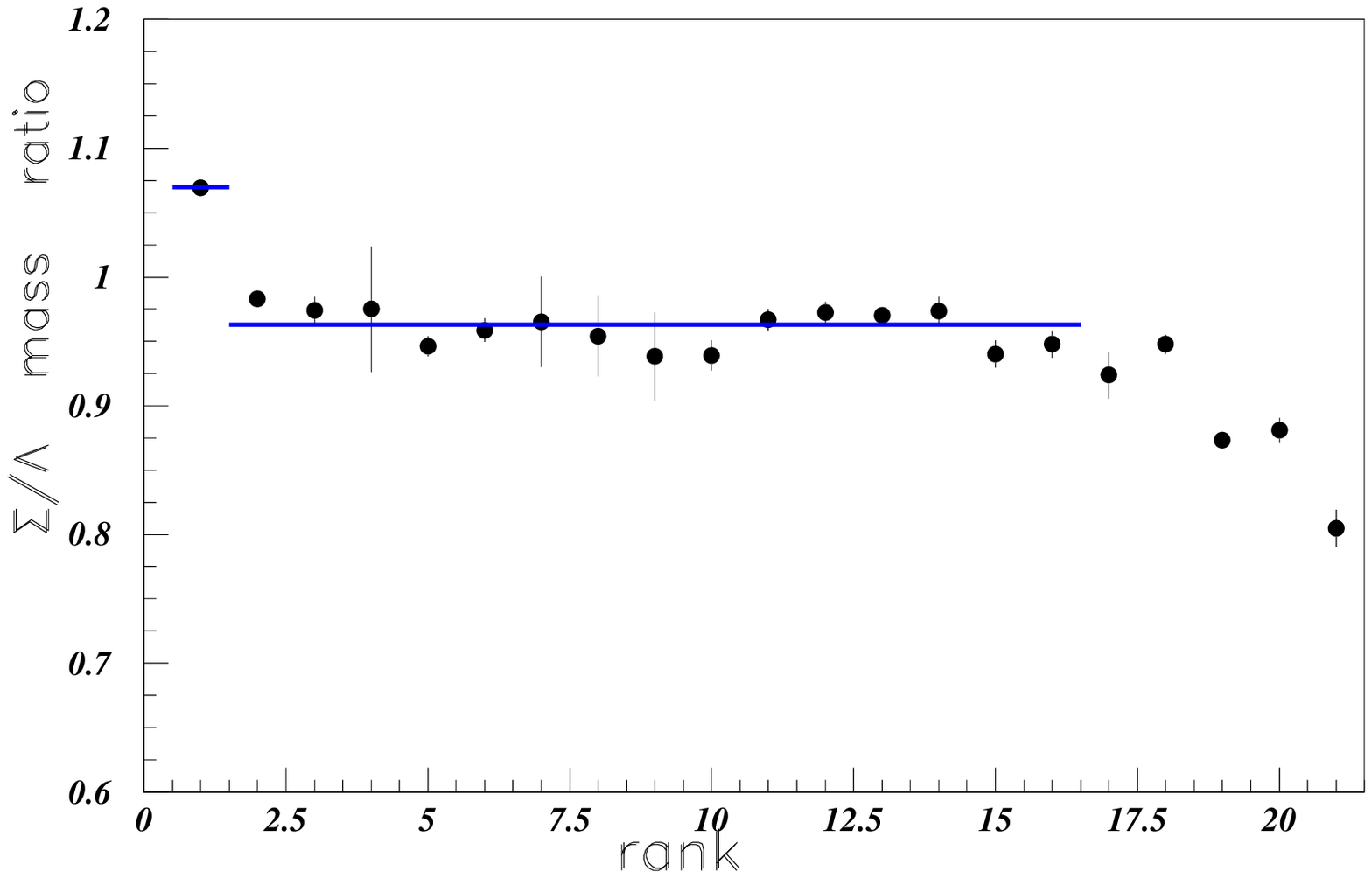}}
\vspace*{-2.mm}
\caption{Black squares show the ratio of baryonic  $\Sigma/\Lambda$ masses. Here, $\Lambda$ data are increased by additionnal masses (see text).}
\end{center}
\end{figure}
\begin{figure}[h]
\begin{center}
\hspace*{-3.mm}
\vspace*{1.mm}
\scalebox{1}[0.9]{
\includegraphics[bb=25 232 515 550,clip,scale=0.45]{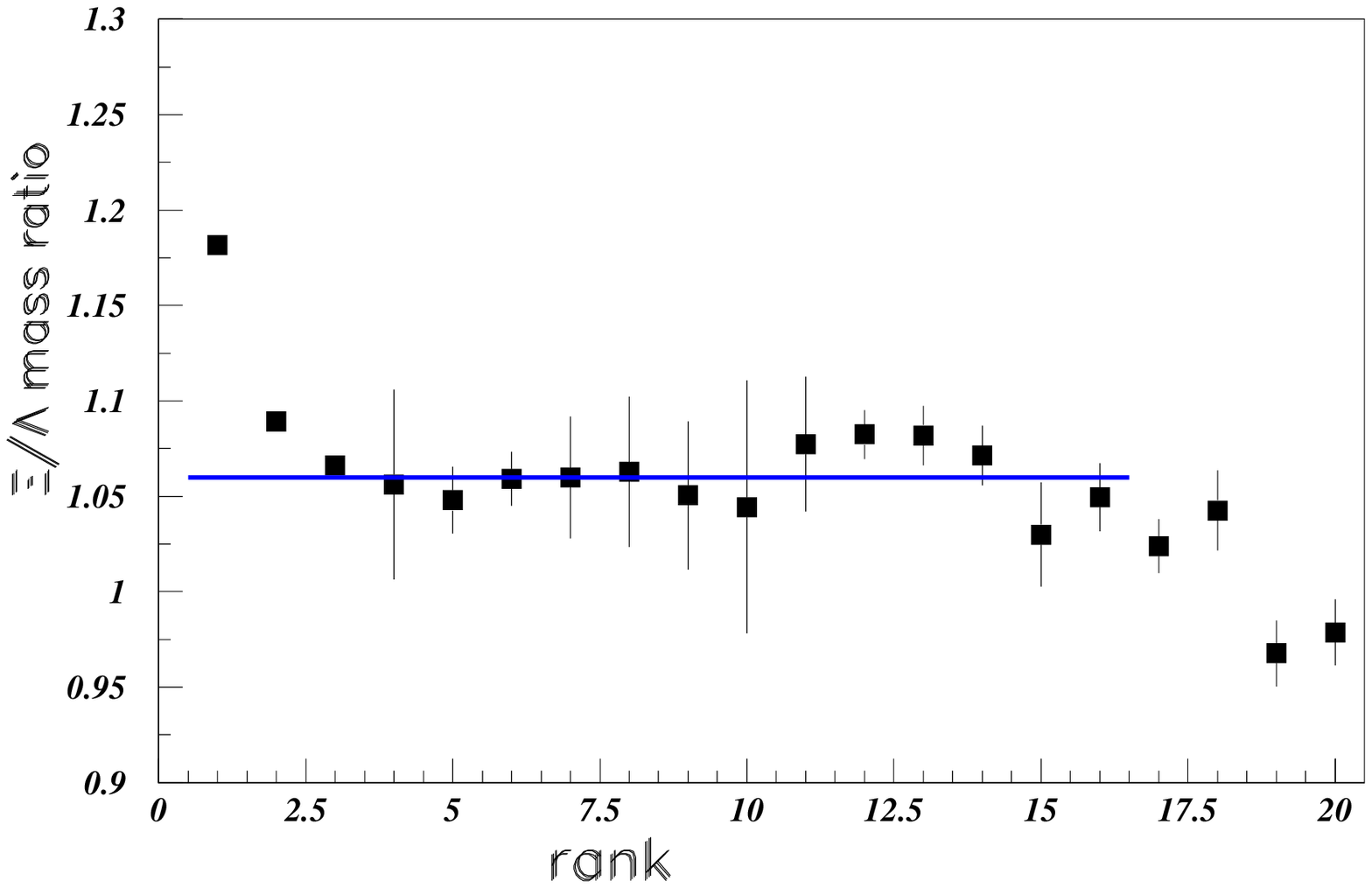}}
\vspace*{-2.mm}
\caption{Black squares show the ratio of baryonic  $\Xi/\Lambda$ masses. Here, $\Xi$ and $\Lambda$ data are increased by additionnal masses (see text).}
\end{center}
\end{figure}
There is also missing masses in the   $\Lambda$ mass table in the range 2110$\le$M~$\le$2325~MeV (see figures 12 and 23), and in the range
2350$\le$M~$\le$2585~MeV (see figure 12). There is no attempt in this paper to suggest other possible $\Lambda$ masses in these ranges.
We observe  in several figures a step between two horizontal lines, again tentatively attributed to the lack of a few $\Lambda$ masses in the range 1700$\le~M~\le$1800~MeV. This step is corrected after the introduction of $\Lambda$ masses at M = 1720~MeV, 1750~MeV, and 1780~MeV. 

We observe that the same mass range was completed by arbitrarily introduced baryons in the three species: N$^{*}$, $\Lambda$, and $\Xi$. Figure~24 shows the $\Xi/\Lambda$ mass ratio, after the introduction of "additionnal masses".
\subsection{Relations giving baryonic mass formula}
The Gell-Mann-Okubo mass formula \cite{okubo} considers the masses of the baryon octet $J^{P}$ = 1/2$^{+}$, and eliminating a few parameters, get the following relation:
\begin{center}
M$_{\Sigma}$ + 3M$_{\Lambda}$ = 2 (M$_{N}$ + M$_{\Xi}$)
\end{center}
The choice of three $\Lambda$ for one $\Sigma$ allows to get a very good precision for the relation. The relative gap between both quantities  of the previous relation is as small as y = 5.6*10$^{-3}$.

In the same spirit, we compute below, a few relations between  yrast masses,
for the same number of quark flavours in both sides of the relation. Restricting ourselves to the choice I = 0 and $J^{+}$ = 1/2$^+$ particles, we observe that the relation:

\begin{center}
M($\Omega^{-}_{b}$) + M($\Lambda^{+}_{c}$) = M($\Omega^{0}_{c}$) + M($\Lambda^{0}_{b}$)
\end{center}
is verified with a precision of 5.0*10$^{-3}$.

In the same way, the addition of the baryonic masses of a $0(1/2+)$ particle 
with a $1/2(1/2+)$ particle, is rather stable,  as shown in the following example where "n" stands for the neutron:  

\begin{center}
M($\Omega^{0}_{c}$) + M(n) = M($\Lambda^{+}$) + M($\Xi^{+}_{c}$)
\end{center}
which is verified with  a precision of 1.4$\%$.

In the same way, the addition of two isospin 1/2 baryonic masses $1/2(1/2+)$ with two isospin 0 baryonic masses $0(1/2+)$ is rather stable, as shown in the following example:

 \begin{center}
M($\Xi^{+}_{c}$) + M($\Xi^{0}_{c}$)  =  M($\Omega^{0}_{c}$) +
 M($\Lambda^{+}_{c}$)
 \end{center}
 which is verified with a precision of 0.86 10$^{-3}$.
 
On the other hand, the equality  is no more nearly obtained if we add the masses of two different and increasing isospin values. For example, the mass of two
I =1 baryonic particles $1(1/2+)$ compared to the sum of two I = 1/2 baryonic particles differs in a larger extend, as shown in the following example:
\begin{center}
M($\Sigma^{++}_{c}$) + M($\Sigma^{0}$) and
M($\Xi^{+}_{c}$) + M(n) 
\end{center}
which differs by 6.8$\%$. The mass of two
I =1 baryonic particles $1(1/2+)$ compared to the sum of a I = 1/2 and I = 3/2 baryonic particles, differs in a smaller extend, as shown in the following example:
\begin{center}
M($\Sigma^{++}_{c}$) + M($\Sigma^{0}$) = M($\Xi^{+}_{c}$) + M($\Delta$) 
\end{center}
which is verified within 1.4 $\%$.

So the precision obtained in the mass computations shown above, can generally be  of the order or better than 1$\%$. For example the mass of the double charmed baryon $\Xi^{++}_{cc}$  = 3.5189~GeV \cite{pdg} can be calculated by the relation: 
M($\Xi^{++}_{cc}$)  = 2 M($\Lambda^{+}_{c}$) - (M(N) + M($\Delta$)) /2 = 3.4874~GeV  and the corresponding relative precision is close to
 y = 9*10$^{-3}$.\\
 In the same way 
M($\Xi^{++}_{cc}$)  =  M($\Lambda^{+}_{c}$)~+~M($\Sigma^{+}_{c}$)~-~M($\Delta$)  is verified within 2.7 10$^{-3}$.

The mass of the $\Omega^{0}_{c}$ = 2695.2~MeV can be calculated by two ways, namely :
$M(\Omega^{0}_{C}$) = 2*$M(\Xi_{C})$ - $M(\Lambda^{+}_{C})$ = 2480.8~MeV, therefore distant by 8.3 $\%$, or by:\\
$M(\Omega^{0}_{C})$ = $M(\Xi^{-}_{C})$ + $M(\Lambda^{+}_{C})$ - M(N) = 2669.6~MeV, therefore distant by 0.95$\%$.
 
When applied again to yrast masses, the relation:
\begin{center}
 M($\Xi_{C}$) + M(N) = M($\Lambda$) + M($\Lambda_{C}$) 
 \end{center}
 is well obtained, since the relative discrepancy between the two members of the equality is equal to $\Delta y(1)/y(1) = 1.6*10^{-3}$. The equality is worse for (not yrast) heavier masses, since $\Delta y(2)/y(2) = 3.9*10^{-3}$, $\Delta y(3)/y(3) = 4.8*10^{-3}$, $\Delta y(4)/y(4) = -9.2*10^{-3}$, $\Delta y(5)/y(5) = -1.9*10^{-2}$, and $\Delta y(6)/y(6) = -5.4*10^{-3}$.

\subsection{Comparison with predictions of theoretical models}
Many models of baryons have been proposed. It is known that "Lattice QCD successfully estimates ground states of hadron spectrum, but excited states still represent an outstanding challenge" \cite{verduci}. The works done within the constituent quark models, before 2000  are reported in \cite{capstick}. 
The model of Isgur and Karl \cite{isgur} predicts the existence of three $N^{*}$ in the mass range 1700$\le~M\le$1720 MeV, and nothing above up to M = 1870~MeV. Concerning the $\Lambda$ masses, this model predicts the existence of a $\Lambda^{+}_{1/2}$ at M = 1740~MeV, and no $\Lambda$ with negative parity, in the mass range 1710$\le~M\le$1800~MeV.  These results do not fit with our "introduced" masses. It is noticeable that the number of calculated baryonic masses is smaller than the number of experimental masses. However "there are far less $\Sigma$ and 
$\Xi$ states established experimentally than expected within quark models" \cite{thomas}. On the contrary, the concept of "missing resonances" was introduced to signify the lack of $N^{*}$ resonances obtained within the quark models, but not observed, and in the same way missing $\Xi^{*}$ states were reported \cite{manley}.

A recent review \cite{huey} reported latest progress with a special interest to heavy and not yet observed baryons. The comparison with experimental masses is difficult for $N^{*}$  since the results of the calculations are presented versus different values of the pion mass squared. Concerning heavier baryons, the calculations predict, below M = 1.8~GeV, only one mass for $\Lambda$ and $\Omega$ (not studied in the present work), and only two masses for $\Sigma$ and $\Xi$ baryons, therefore a much lower number than experimentally observed.

There is an often discussed problem about nucleon states, predicted by quark models, but not observed experimentally. These states are called "missing resonances", already mentionned. A search for these resonances via associated strangeness photoproduction was performed
\cite{saghai}. The authors concluded that two such resonances, namely a $S_{11}$, M = 1820~MeV, and a $D_{13}$, M = 1920~MeV, help to improve the  cross-section and the recoil polarization asymmetry between calculation and experimental data studied in the reaction $\gamma$p $\to$ $K^{+}\Lambda$. Another calculation was performed to study the possible improvement of the cross-sections and polarization asymmetries by adding several (S11, P11, P13, D13, D15, and H1,11) new resonances to all PDG baryonic resonances \cite{he}. These authors developped a chiral quark model for the $\gamma$p $\to$ $\eta$p reaction.
In their model A, they introduced six new resonances, but only one $S_{11}$(1730) happens to play a significant role. 

A systematic search was undertaken at CLAS \cite{clas} \cite{crede}, using photon beams in the range 0.8$\le~E_{\gamma}\le$4.0~GeV/c and looking to different disintegration channels. Another study was performed by the CBELSA/TAPS experiment at ELSA \cite{thoma}, studying also photoproduction of single and multi-meson final states off the nucleon, and enriched by double polarization experiments using circularly or linearly polarized photon beams \cite{schmidt} \cite{beck}. Recent measurements were also done at MAMI \cite{schumann}. The aim of these measurements is to be able to construct unambiguously the scattering amplitudes. The existence of several broad overlapping resonances in the mass region 1500$\le~M~\le$2000~MeV, difficult to disentangle, explains that there is, up to now, no clear experimental result concerning these "missing resonances".

\begin{figure}[h]
\begin{center}
\hspace*{-3.mm}
\vspace*{1.mm}
\scalebox{1}[0.9]{
\includegraphics[bb=38 240 530 540,clip,scale=0.45]{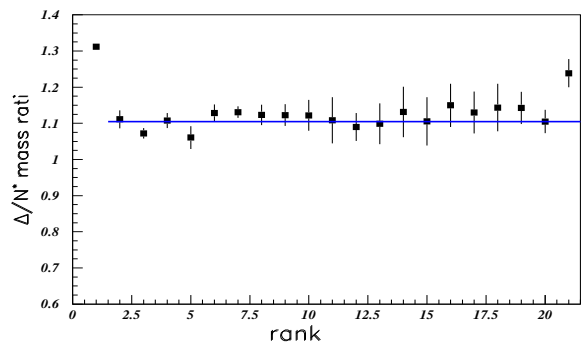}}
\vspace*{-2.mm}
\caption{Black squares show the ratio of baryonic $\Delta/N^{*}$ masses.
The PDG $N^{*}$ data are increasd by additionnal masses (see text).}
\end{center}
\end{figure}
\begin{figure}[h]
\begin{center}
\hspace*{-3.mm}
\vspace*{1.mm}
\scalebox{1}[0.9]{
\includegraphics[bb=38 240 530 540,clip,scale=0.45]{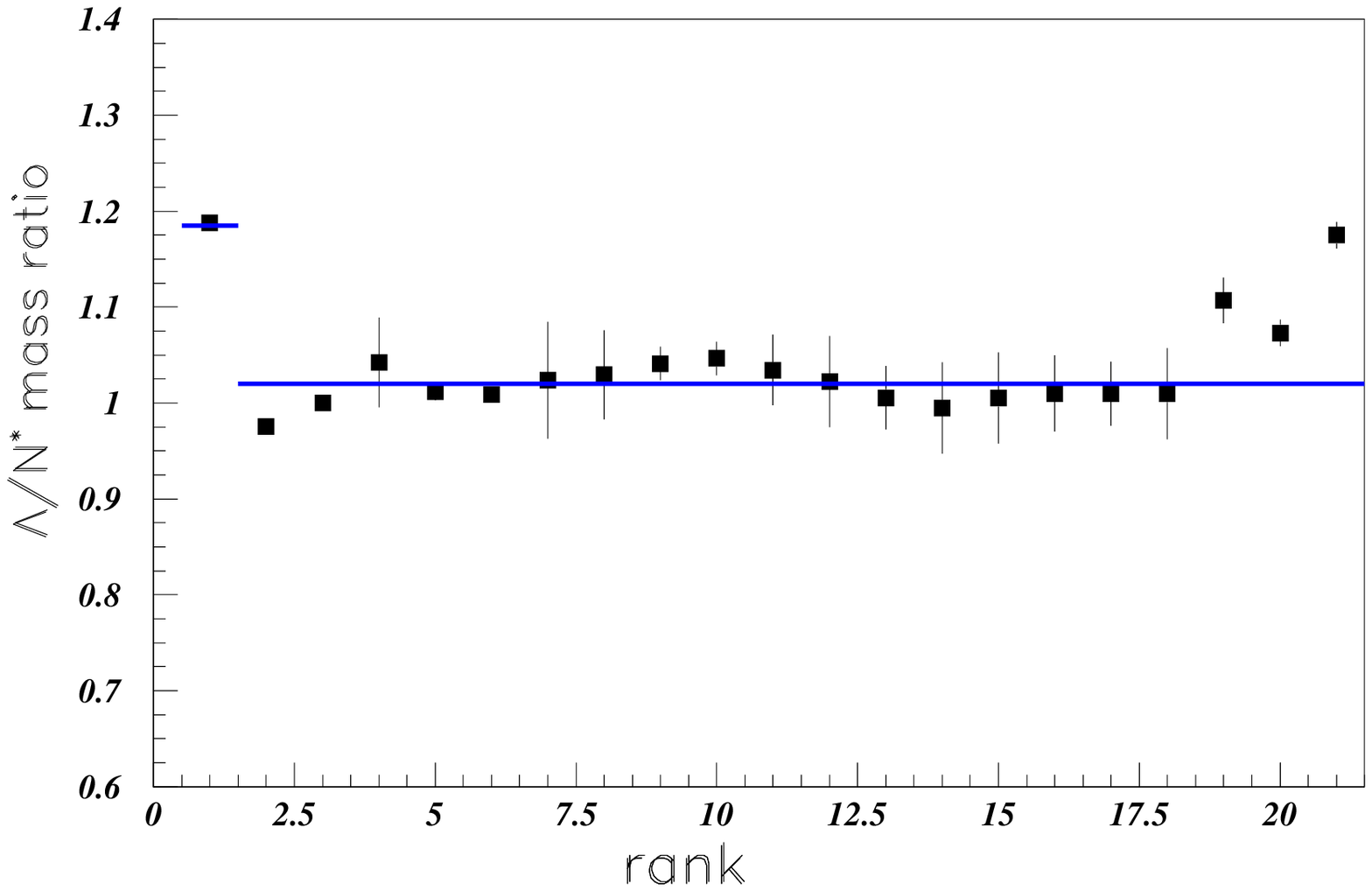}}
\vspace*{-2.mm}
\caption{Black squares show the ratio of baryonic $\Lambda/N^{*}$ masses. The PDG $N^{*}$ and $\Lambda$ data are increasd by additionnal masses (see text).}
\end{center}
\end{figure}
\begin{figure}[h]
\begin{center}
\hspace*{-3.mm}
\vspace*{1.mm}
\scalebox{1}[0.9]{
\includegraphics[bb=38 240 530 540,clip,scale=0.45]{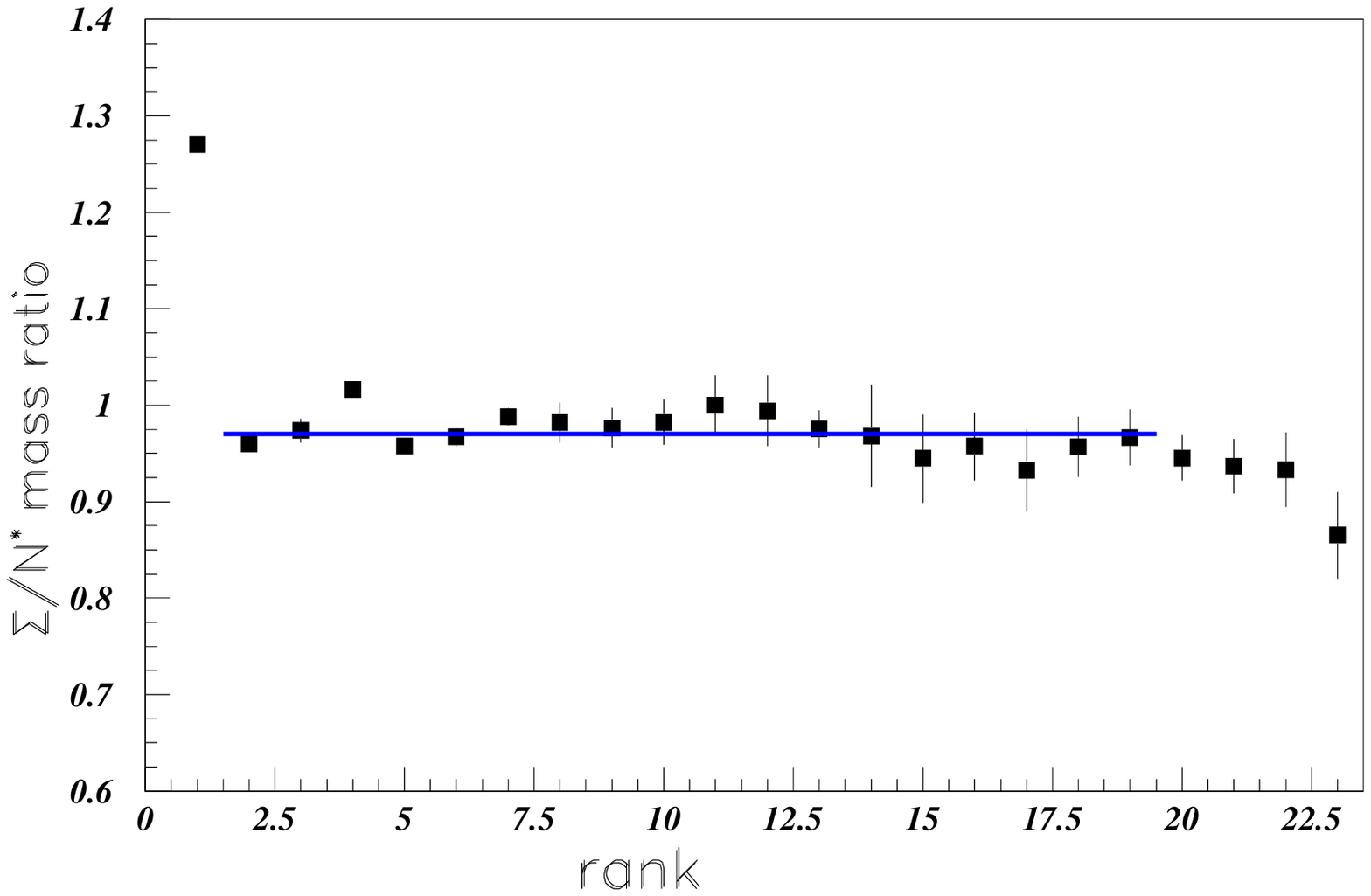}}
\vspace*{-2.mm}
\caption{Black squares show the ratio of baryonic  $\Sigma/N^{*}$ masses.
The PDG $N^{*}$ data are increasd by additionnal masses (see text).}
\end{center}
\end{figure}
\begin{figure}[h]
\begin{center}
\hspace*{-3.mm}
\vspace*{1.mm}
\scalebox{1}[0.9]{
\includegraphics[bb=38 240 530 540,clip,scale=0.45]{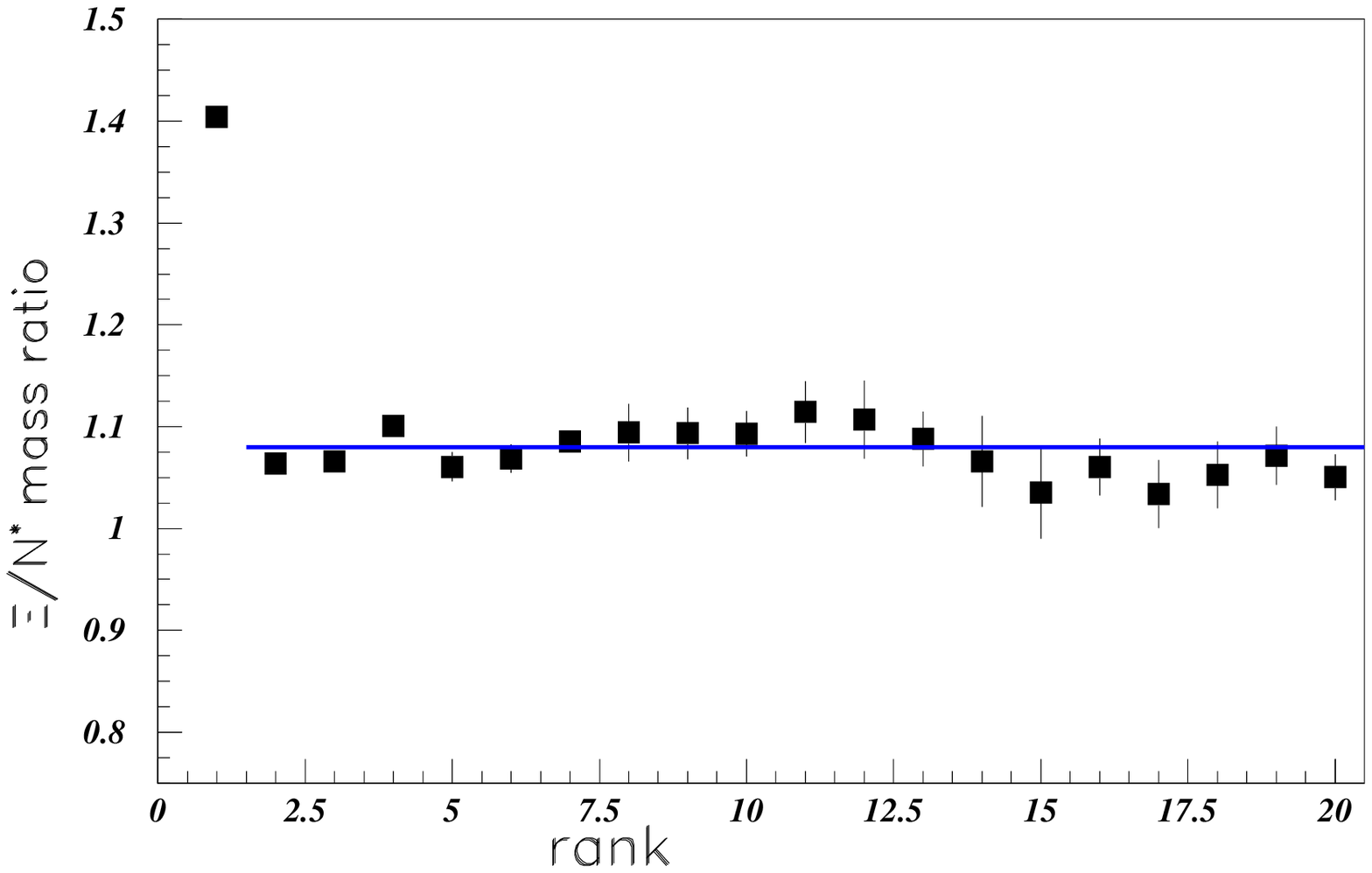}}
\vspace*{-2.mm}
\caption{Black squares show the ratio of baryonic  $\Xi/N^{*}$ masses. The PDG $\Xi$ and $N^{*}$ data are increasd by additionnal masses (see text). }
\end{center}
\end{figure}

\subsection{Discussion concerning the mass ratios between different baryonic species}
Table~II shows the ratios of  baryonic masses between different species, when the first (yrast) mass of all species is ignored. These ratio values are read from the previous figures (2 - 28) in the following way: M($\Lambda$) / M($\Delta$) = 0.94.
The ratio of $\Lambda^{+}_{C}\Sigma^{++}_{C}$ excited baryonic masses over all other excited baryonic masses is rather stable. The same property is observed for $\Xi_{C}$ excited baryonic masses (except for the ratio
$\Xi_{C}/\Lambda^{+}_{C}\Sigma^{++}_{C}$). The substitution of a quark "s" to a quark "u" or "d", does not increase substancially the masses. Moreover, the baryonic masses containing a quark "c" are clearly heavier.

As already mentionned,  these ratios are different for the first mass of all baryonic species. Figure 1 shows that the gap between the excited state spectra and the first (yrast) mass, decreases for increasing baryonic masses. Table~III shows the ratio 
of the first (yrast) mass between two baryonic species. 
The numbers are larger than in table~II, but still exhibit some regularity. The $\Delta$ masses are larger than expected from the continuity due to increasing flavour;  this effect is induced by  larger isospin I = 3/2. The effect of smaller isospins is negligible. 
 However the ratio between two ratios is about the same, for yrast masses (table~III) and excited level masses (table~II). For example the ratio between masses, symbolized by their quark content: [qsc]/[qqc], equals 1.065 for excited levels (table~II) and 1.08 for yrast levels (table~III). The ratio [qss]/[uus] equals 1.107 for excited levels (table~II) and 1.105 for yrast levels (table~III).

\section{Conclusion}
The masses of the different baryonic species are compared, without consideration of their corresponding widths, neither their quantum numbers, or their disintegration channels. In the same way, nothing is taken into account about the possiblity that these baryons could be hybrid (with a part of glue).

Using similar simple ideas, the baryonic fundamental or excited masses have already been discussed \cite{capstick} \cite{valcarce}. However none of these papers has shown the constant ratios between  masses of different baryonic species.

Here, proportionality between all baryonic species masses, except the yrast ones, is observed. It allows  to tentatively predict several missing masses in some species. These predictions are validated by the clear association between gaps in the mass ratios, and large empty mass ranges. An unique set of additionnal mass data is able to correct the shapes of the ratios which do not follow the proportionnality. Three missing masses, in the range 1720$\le$~M~$\le$1820~MeV, are tentatively introduced in the $N^{*}$ and $\Lambda$ tables. In addition to two masses, introduced in the $\Xi$ baryon mass table in the same range, twelve masses are tentatively introduced in this $\Xi$ baryon mass table at higher masses.   

This observation of proportionnality, indicates that common laws apply to get excited state masses of three quarks. The various contributions to the excited state masses, due to different quantum numbers, give rise to small contributions to masses.

\begin{table*}[h]
\begin{center}
\caption{Color on line. Baryonic masses used. \\
PDG masses (black) and masses tentatively \\
introduced (red). (See text).}
\label{Table I}
\begin{tabular}[t]{c c c}
\hline
\hspace*{0.mm}Baryons&masses  (in MeV)\\
\hline
\hspace*{0.mm}N&939, 1440, 1520, 1535, 1655, 1675, 1685\\
\hspace*{0.mm}&1700, 1710, 1720, \textcolor{red}{1750, 1780, 1820}, 1900\\
\hspace*{0.mm}&1990, 2000, 2080, 2090, 2100, 2190, 2200\\
\hspace*{0.mm}&2220, 2275, 2600, 2700, 3000\\
\hspace*{0.mm}$\Delta$&1232, 1600, 1630, 1700, 1750, 1890, 1900\\
\hspace*{0.mm}&1910, 1920, 1930, 1940, 1940, 2000, 2150\\
\hspace*{0.mm}&2200, 2300, 2350, 2390, 2400, 2420, 2750\\
\hspace*{0.mm}&2950\\
\hspace*{0.mm}$\Lambda$,&1115.68, 1405, 1520, 1600, 1670, 1690, \textcolor{red}{1720}\\
\hspace*{0.mm}&\textcolor{red}{1750, 1780}, 1800, 1810, 1820, 1830, 1890\\
\hspace*{0.mm}&2000, 2020, 2100, 2110, 2325, 2350, 2585\\
\hspace*{0.mm}$\Sigma$&1193, 1385, 1480, 1560, 1580, 1620\\
\hspace*{0.mm}&1660, 1670, 1690, 1750, 1770, 1775, 1840\\
\hspace*{0.mm}&1880, 1915, 1940, 2000, 2030, 2070, 2080\\
\hspace*{0.mm}&2100, 2250, 2455, 2620, 3000, 3170\\
\hspace*{0.mm}$\Xi$&1318.28, 1531.8, 1620, 1690, \textcolor{red}{1750, 1790}\\ 
\hspace*{0.mm}&1823, \textcolor{red}{1860, 1870, 1880}, 1950, \textcolor{red}{1970, 1980}\\
\hspace*{0.mm}&2025, \textcolor{red}{2060}, 2120, \textcolor{red}{2150, 2200}, 2250\\
\hspace*{0.mm}&\textcolor{red}{2300, 2310, 2340, 2360}, 2370, 2500\\
\hspace*{0.mm}$\Lambda^{+}_{C}$,$\Sigma^{++}_{C}$&2286.46, 2453.76, 2518.4, 2595.4\\
\hspace*{0.mm}&2628.1, 2766.6, 2802, 2881.53, 2939.3\\
\hspace*{0.mm}$\Xi_{C}$&2469.5, 2577.8, 2645.8, 2789.2, 2817.4\\
\hspace*{0.mm}&2931, 2974, 3054.2, 3077, 3122.9\\
\hline
\end{tabular}
\end{center}
\end{table*}

\samepage

\begin{table*}[!h]
\begin{center}
\caption[table]{Ratios between masses of two different baryonic species, the first (yrast) mass of each species 
being ignored.  $\Lambda^{+}_{C}$ means implicitely  $\Lambda^{+}_{C}$ and $\Sigma^{++}_{C}$.}
\label{Table II}
\begin{tabular}{c c c c c c c c c}
\hline
\hspace*{0.mm} &N$^{*}$&$\Delta$&$\Lambda$&$\Sigma$&$\Xi$&$\Omega$&$\Lambda_{c}^{+}$&$\Xi_{c}$\\
 &udq&uuu&uds&uus&qss&sss&qqc&qsc\\
\hline
\hspace*{0.mm}N$^{*}$&1&1.105&1.02&0.97&1.08&$\bullet$&1.65&1.77\\
\hspace*{0.mm}$\Delta$&$\bullet$&1&0.94&0.884&0.99&$\bullet$&1.6&1.6\\
\hspace*{0.mm}$\Lambda$&$\bullet$&$\bullet$&1&0.963&1.06&$\bullet$&1.6&1.705\\
\hspace*{0.mm}$\Sigma$&$\bullet$&$\bullet$&$\bullet$&1&1.107&$\bullet$&1.685&1.815\\
\hspace*{0.mm}$\Xi$&$\bullet$&$\bullet$&$\bullet$&$\bullet$&1&$\bullet$&1.54&1.64\\
\hspace*{0.mm}$\Omega$&$\bullet$&$\bullet$&$\bullet$&$\bullet$&$\bullet$&1&$\bullet$&$\bullet$\\
\hspace*{0.mm}$\Lambda_{c}^{+}$&$\bullet$&$\bullet$&$\bullet$&$\bullet$&$\bullet$&$\bullet$&1&1.065\\
\hspace*{0.mm}$\Xi_{c}$&$\bullet$&$\bullet$&$\bullet$&$\bullet$&$\bullet$&$\bullet$&$\bullet$&1\\
\hline
\end{tabular}
\end{center}
\end{table*}

\samepage
\nopagebreak[4]

\begin{table*}[!h]
\begin{center}
\caption[table]{Ratios between the first (yrast) mass of\\
two different baryonic species. $\Lambda^{+}_{C}$ means implicitely\\  $\Lambda^{+}_{C}$ and $\Sigma^{++}_{C}$.}
\label{Table III}
\begin{tabular}{c c c c c c c c c}
\hline
\hspace*{0.mm} &N$^{*}$&$\Delta$&$\Lambda$&$\Sigma$&$\Xi$&$\Omega$&$\Lambda_{c}^{+}$&$\Xi_{c}$\\
\hspace*{0.mm} &udq&uuu&uds&uus&qss&sss&qqc&qsc\\
\hline
\hspace*{0.mm}N$^{*}$&1&1.312&1.188&1.27&1.404&1.781&2.435&2.63\\
\hspace*{0.mm}$\Delta$&$\bullet$&1&.9056&0.9685&1.07&1.358&1.86&2.00\\
\hspace*{0.mm}$\Lambda$&$\bullet$&$\bullet$&1&1.07&1.06&1 .50&2.05&2.21\\
\hspace*{0.mm}$\Sigma$&$\bullet$&$\bullet$&$\bullet$&1&1.105&1.40&1.92&2.07\\
\hspace*{0.mm}$\Xi$&$\bullet$&$\bullet$&$\bullet$&$\bullet$&1&1.269&1.73&1.87\\
\hspace*{0.mm}$\Omega$&$\bullet$&$\bullet$&$\bullet$&$\bullet$&$\bullet$&1&1.367&1.467\\
\hspace*{0.mm}$\Lambda_{c}^{+}$&$\bullet$&$\bullet$&$\bullet$&$\bullet$&$\bullet$&$\bullet$&1&1.08\\
\hspace*{0.mm}$\Xi_{c}$&$\bullet$&$\bullet$&$\bullet$&$\bullet$&$\bullet$&$\bullet$&$\bullet$&1\\
\hline
\end{tabular}
\end{center}
\end{table*}

\end{document}